# Quasi-Simultaneous Broadband Spectral Energy Distributions of a Sample of Fermi Blazars - I. Correlation Results

Yi Zhong,[1] Zhujian Wan,[2,3] Rui Xue[1]★ Hubing Xiao,[4]† Dingrong Xiong,[5,6,7]‡ and Ze-Rui Wang[8,9]
[1]*Department of Physics, Zhejiang Normal University, Jinhua 321004, China*
[2]*Key Laboratory of Dark Matter and Space Astronomy, Purple Mountain Observatory, Chinese Academy of Sciences, Nanjing 210023, People's Republic of China*
[3]*School of Astronomy and Space Science, University of Science and Technology of China, Hefei, Anhui 230026, People's Republic of China*
[4]*Shanghai Key Lab for Astrophysics, Shanghai Normal University, Shanghai 200234, China*
[5]*Yunnan Observatories, Chinese Academy of Sciences, 396 Yangfangwang, Guandu District, Kunming, 650216, People's Republic of China*
[6]*Center for Astronomical Mega-Science, Chinese Academy of Sciences, 20A Datun Road, Chaoyang District, Beijing, 100012, People's Republic of China*
[7]*Key Laboratory for the Structure and Evolution of Celestial Objects, Chinese Academy of Sciences, 396 Yangfangwang, Guandu District, Kunming, 650216, People's Republic of China*
[8]*College of Physics and Electronic Engineering, Qilu Normal University, Jinan 250200, China*
[9]*Shandong Key Laboratory of Space Environment and Exploration Technology, China*

**ABSTRACT**

Blazars' non-thermal emission shows rapid variability across all wavelengths, so spectral energy distributions (SEDs) built from quasi-simultaneous data are crucial for revealing the jets physical properties. In this work, we construct quasi-simultaneous broadband SEDs for 93 Fermi blazars (56 FSRQs, 35 BL Lacs, and 2 blazar candidates of uncertain type), fit both peaks with cubic functions to allow for potential asymmetries, and examine correlations among key parameters. Our main results are summarized as follows: (1) We find that synchrotron peak frequency and curvature are only weakly related, suggesting that charged particles are accelerated by mixed acceleration mechanism. (2) The blazar sequence is confirmed in the observer's frame through negative correlations of both the bolometic luminosity $\log L_{\rm bol}$ and the Compton dominance $\log Y$ with the synchrotron peak frequency $\log \nu_{\rm syn}^{\rm peak}$. After correcting for Doppler boosting, a weak positive correlation emerges between $\log L_{\rm bol}$ and $\log \nu_{\rm syn}^{\rm peak}$. FSRQs and BL Lacs exhibit distinct correlation patterns within the blazar sequence, indicating differences in cooling mechanisms. (3) Using variability time lags between 0.1–1 GeV and 1–300 GeV light curves, we estimate lower limits of Doppler factors for 4 sources, providing a jet-speed diagnostic anchored directly to the $\gamma$-ray emission zone.

**Key words:** radiation mechanisms: non-thermal – galaxies: active – galaxies: jets.

## 1 INTRODUCTION

Blazars are a special subclass of active galactic nuclei (AGNs) with jets pointed toward the observer. Based on the equivalent width (EW) of their emission lines, blazars are divided into flat-spectrum radio quasars (FSRQs), which exhibit strong broad emission lines (EW $\geq$ 5 Å), and BL Lacertae objects (BL Lacs), displaying weak or no emission lines (EW < 5 Å; Urry & Padovani 1995). Both of them emit intense radiation across the entire electromagnetic spectrum, from radio to gamma-ray bands.

An important tool to describe the characteristics of blazars is the spectral energy distribution (SED), which displays a double-bump structure (Abdo et al. 2010b). The formation of this double-bump structure is closely related to the radiation mechanisms of relativistic charged particles in the jet. In leptonic models, it is generally believed that the low-energy peak is attributed to synchrotron radiation from relativistic electrons in the jet, while the high-energy peak is attributed to inverse Compton (IC) scattering (Massaro et al. 2004; Meyer et al. 2012). However, there remains controversy regarding the origin of the soft photons involved in IC scattering. The first scenario suggests that these photons originate from synchrotron radiation, known as the synchrotron self-Compton (SSC) process (Rees 1967; Jones et al. 1974; Marscher & Gear 1985; Maraschi et al. 1992; Sikora et al. 1994; Bloom & Marscher 1996; Tavecchio et al. 1998), while the second suggests they come from external regions outside the jet, known as the external Compton (EC) process (Dermer et al. 1992; Dermer & Schlickeiser 1993; Ghisellini & Tavecchio 2009). The hadronic model is also widely used to explain the high-energy peak (e.g., Aharonian 2000; Mücke & Protheroe 2001; Böttcher et al. 2013; Xue et al. 2022, 2023; Wang et al. 2024). While the SEDs of blazars remain in low states during most of the time, they can stochastically exhibit flaring states. Due to their rapid multi-wavelength variabilities, SEDs constructed from historical data cannot completely describe the basic properties in different flaring periods. Giommi et al. (2012) compared radio data from Planck and WMAP, X-ray data from Swift and BZCAT, as well as gamma-ray data observed by *Fermi*-LAT in both short-term and long-term periods, and found that the uncertainties associated with non-simultaneous radio data are relatively minor, whereas non-simultaneous high-energy band data show significantly larger uncertainties (by a factor of ten or more). This suggests that using non-simultaneous data may result in misinterpretations of the SED shape, and both spectral peaks, especially the high-energy peak with significant shifts over time. Consequently, multi-wavelength SEDs built from quasi-simultaneous data are necessary to accurately reveal their physical properties.

★ E-mail: ruixue@zjnu.edu.cn
† E-mail: hubing.xiao@shnu.edu.cn
‡ E-mail: xiongdingrong@ynao.ac.cn

The construction of SED models is typically categorized according to their radiation mechanisms and physical origins. The main theoretical frameworks include leptonic models, hadronic models, one-zone radiation models, and multi-zone radiation models. To distinguish these models, matching key physical parameters with observational parameters is required. For instance, the relationship between the curvature and the peak frequency helps determine particle acceleration mechanisms (Massaro et al. 2004; Paggi et al. 2009; Rani et al. 2011; Tramacere et al. 2007, 2009, 2011; Chen 2014; Xiao et al. 2024b, 2025). Based on the bolometric luminosity, Compton dominance, and the peak frequency of the synchrotron peak, it can help to study the physical properties of the jet implied by the blazar sequence (Fossati et al. 1998; Ghisellini et al. 1998; Wan et al. 2024). Furthermore, it is also possible to determine the location, radius, and Doppler factor of the emission region based on the multiwavelength variability time-scale (e.g., Hu et al. 2024). X-ray polarization degree serves as a critical diagnostic tool to distinguish leptonic models and hadronic models (e.g., Zhang et al. 2024b).

When studying the properties of blazar jets through SEDs, empirical functions offer a model-independent way to characterize their shapes. Parameters derived from such fits, including peak frequencies, spectral curvatures, and peak luminosities, serve as useful diagnostics of jet physics (e.g., acceleration mechanisms; Massaro et al. 2004; Paggi et al. 2009; Chen 2014). These parameters can further be used to constrain the physical parameters adopted in theoretical models (Tavecchio et al. 1998; Hu et al. 2024). Among the commonly used empirical functions, quadratic (Chen 2014; Fan et al. 2016b; Yang et al. 2022a, 2023) and cubic functions (Abdo et al. 2010a) are generally employed. The quadratic function is axially symmetric, whereas observations show that the SEDs of blazars, particularly those with high synchrotron peak frequencies, display asymmetric structures (Xue et al. 2016). In such cases, fitting with a quadratic function would introduce systematic bias in the derived peak parameters. Recently, Deng et al. (2024) compare the fitting results with the quadratic and cubic functions using the Akaike Information Criterion Corrected (AICc), and find that the cubic function provides a better fit for asymmetric SEDs, whereas for symmetric SEDs the two functions yield comparable results. By reducing bias in estimates of peak parameters, cubic fitting yields more reliable observational parameters. These improved measurements enhance correlation analyses and help place stronger constraints on jet properties. The construction of multi-wavelength SEDs, especially quasi-simultaneous ones, requires the inclusion of $\gamma$-ray data. The *Fermi*-LAT Fourth Data Release (DR4) has expanded the number of blazars with detected $\gamma$-ray emission, thereby identifying a larger set of sources for which quasi-simultaneous data can be assembled from other archives and observations. In previous studies, quasi-simultaneous SEDs are usually built for specific objects in flare states (e.g., Abdalla et al. 2020; Acciari et al. 2020; Liu et al. 2025), and most statistical analyses have relied on historical data (e.g., Fan et al. 2016b; Yang et al. 2022a, 2023, cf., Abdo et al. (2010b); Giommi et al. (2012)), which may introduce significant biases into the results. In this work, we construct a substantially large sample of blazars with quasi-simultaneous SEDs and fit them with cubic functions, which better capture the asymmetric shapes of blazar spectra (e.g., Xue et al. 2016; Deng et al. 2024), and reduce systematic errors in peak parameters. This approach provides more reliable observational quantities, strengthens correlation analyses, and discuss their physical implications. The structure of this paper is as follows. In Section 2, we describe the sample selection and gamma-ray data reduction. Section 3 presents the SED fitting results. Section 4 provides the correlations results and discussions. Finally, we draw conclusions in Section 5. The cosmological parameters $H_0 = 70\ \mathrm{km\ s^{-1} Mpc^{-1}}, \Omega_\mathrm{M} = 0.27, \Omega_\Lambda = 0.73$ are used in this work (Bennett et al. 2014).

## 2 SAMPLE COLLECTION AND DATA ANALYSIS

Our samples are selected from all blazar subclasses , including FSRQs, BL Lacs, and blazar candidates of uncertain type (BCUs),in the 4FGL-DR4 catalog[1]. The construction of quasi-simultaneous broadband SEDs for Fermi blazars critically depends on the availability of multi-wavelength data outside the $\gamma$-ray band. To this end, we select sources with measured redshift and extract their simultaneous observational data from infrared to X-ray bands using the Science Data Center (SSDC) SED Builder[2] (Stratta et al. 2011). To meet the requirements for simultaneous multi-wavelength observations, the time intervals for the data we collected is within seven days as suggested by Giommi et al. (2012). It should be noted that some available data in the SSDC span quite a longe interval, which does not meet the requirement of simultaneous multi-wavelength observations. Taking the first source in our sample, i.e., PKS 0019+058, as an example, some of its X-ray data (e.g., with start time $T_\mathrm{Start}$ = 54968.05 and stop time $T_\mathrm{Stop}$ = 55885.52, given in MJD) fall outside the defined time window and are therefore excluded. In screening the data, it is required that all selected data should fall entirely within the designated period. For instance, some ultraviolet observations (e.g., $T_\mathrm{Start}$ = 54973.48, $T_\mathrm{Stop}$ = 54973.48) are included as they lie entirely within the window. Similarly, X-ray data (e.g., $T_\mathrm{Start}$ = 54968.05, $T_\mathrm{Stop}$ = 54698.06) that satisfy this criterion are also selected. After filtering, the sample has 93 blazars, including 56 FSRQs, 35 BL Lacs and 2 BCUs.

Note that the radio data from SSDC were mostly collected in the 1990s and show larger time gaps than those in other bands, so they are excluded during our data screening. Infrared data is crucial for constraining the rising side of the low-energy peak; however, in some cases the available infrared data do not satisfy the requirement of simultaneous observations. In such situations, we adopt radio data above the turnover frequency to replace infrared data in constraining the low-energy peak, following theoretical studies which suggest that emission above the turnover frequency can reasonably be assumed to originate from the same optically thin region, whereas radio emission below the turnover frequency is produced by the superposition of multiple optically thick regions (Ghisellini et al. 2010; Potter & Cotter 2012; Liu et al. 2023). Our data selection criteria are based solely on data simultaneity and therefore cannot distinguish between flaring and quiescent states. This limitation is not unique to X-ray data, as optical observations may also encounter similar challenges. Radio data are less problematic in this regard because their variability is generally much weaker than in the optical and X-ray bands, with typical timescales of months to years. For this reason, the use of archival high-frequency radio data results in minor spectral distortions (Giommi et al. 2012). We acknowledge that a seven-day window may contain data from different activity states, which may

[1] https://fermi.gsfc.nasa.gov/ssc/data/access/lat/
[2] https://tools.ssdc.asi.it/SED/

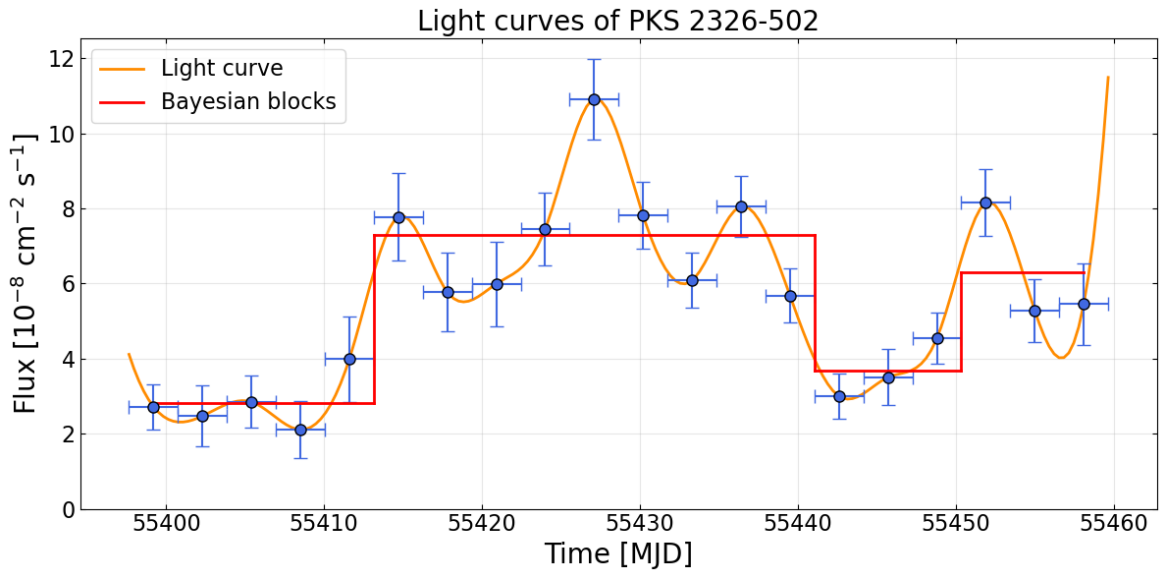

**Figure 1.** Light curves of sources exhibiting $\gamma$-ray variabilities. Light curves are generated using either constant-binning or adaptive-binning methods, depending on whether any time bin in the initial constant-binning light curve satisfy $TS > 100$. Gold curve shows the spline that highlights the primary trend of variation, and red curve shows the Bayesian blocks.

(The complete figure set (13 images) is available.)

introduce bias. After careful inspection of the SEDs, we find that clear multi-state cases are rare, and thus the overall influence on our results should be minor. Although the SSDC database does not provide variability information for infrared–X-ray data, we use the contemporaneous Fermi-LAT data as an indirect indicator of the source activity state within the selected simultaneous windows, as described below, while noting that this does not fully ensure matching activity states across all bands.

In the SSDC database, Fermi-large Area Telescope (Fermi-LAT) $\gamma$-ray data are collected by annual integration, which cannot satisfy the requirement of quasi-simultaneous observations. Because the number of $\gamma$-ray photons is much smaller than in lower-energy bands, useful spectra can be obtained during strong flares or by integrating over longer timescales. To ensure sufficient photon statistics, we therefore adopt a two-month integration window for the Fermi-LAT data, extending one month forward and backward around the simultaneous time period of the lower-energy bands. This approach has been widely employed in previous studies and is generally regarded as fulfilling the requirement of quasi-simultaneous observations (Abdo et al. 2010b; Giommi et al. 2012), while minimizing the impact of temporal averaging on the physical interpretation of the SEDs.

We download the photon and spacecraft files from the public Fermi Science Support Center (FSSC)[3]. The region of interest centered on the right ascension (RA) and declination (DEC) of each source is set to 15°. To prevent the point source analysis from being challenging, we follow the standard data selection recommendations and restrict the photon energy range from 100 MeV to 300 GeV. After obtaining the data, we analyze it using *easyfermi*[4] (de Menezes 2022), which is a graphical user interface (GUI) tool developed based on *Fermipy*[5] (Wood et al. 2017). The analysis employs the standard Galactic diffuse emission model (`gll_iem_v07.fits`) and isotropic background model (`iso_P8R3_SOURCE_V3_v1.txt`). When generating the energy spectrum, we dynamically adjust the number of energy bins based on the Test Statistic ($TS$), initially setting 6 bins. Only bins with $TS > 9$ are retained as valid data points. Bins with $TS \leq 9$ or flux errors exceeding half of the flux value are marked as 95% confidence-level flux upper limits. If all 6 initial bins result in upper limits, we reduce the number of bins to 3 to enhance photon statistics in individual energy bands. We also use *easyfermi* to generate the $\gamma$-ray light curve of the two-month period. The system invokes the `lightcurve()` function from *Fermipy* to produce a constant-binning result with 20 time bins, each corresponding to three days. Within the same procedure, *easyfermi* also applies the adaptive-binning method (Lott et al. 2012), whenever the $TS$ of a bin exceeds a specified threshold, to identify variability occurring on timescales shorter than the three-day bin in the constant-binning analysis. This adaptive-binning method adjusts the bin width to achieve a constant relative flux uncertainty, thereby improving time resolution during periods of enhanced activities. By specifying the $TS$ threshold ($TS_{\rm Threshold}$) and the number of iterations ($N^{\circ}$ iterations) in the graphical interface, any interval with $TS > 2 \times TS_{\rm Threshold}$ is subdivided into smaller time intervals. Here we set $TS_{\rm Threshold} = 50.0$ and the number of iterations $N = 7$. Then we further use Bayesian block method[6] to search for intervals where the flux changes by more than a factor of two, which we regard as variability events (Scargle et al. 2013; Krauß et al. 2016). Following these steps, we obtain $\gamma$-ray light curves for all sources. The Bayesian block method reveals that 10 out of 56 FSRQs and 3 out of 35 BL Lacs exhibit variability in their $\gamma$-ray light curves, and their light curves are presented in Figure 1. For all 13 variable sources, we derive the minimum $\gamma$-ray variability timescale, $\tau_{\rm var}$, from the adaptive-binning light curves, where $\tau_{\rm var}$ is defined as the smallest bin width reached during the iterative process. The fact that only 13 out of 93 sources show significant $\gamma$-ray variability within our two-month window, and that most of them vary only by a factor of ~2–3, suggests that our quasi-simultaneous SEDs are not predominantly assembled during strong $\gamma$-ray flaring episodes. Nevertheless, we emphasize that residual state mismatches between the $\gamma$-ray and lower-energy bands cannot be fully excluded due to the lack of multi-band variability information; such effects may increase the scatter and contribute to the systematic uncertainties of the derived correlations. Using $\tau_{\rm var}$ enables us to follow the method of Hu et al. (2024) for estimating the Doppler factor $\delta$ in the $\gamma$-ray emitting region (see Section 4.5), and provides valuable constraints on the size of the emitting region ($R \leqslant \delta\tau_{\rm var}c/(1+z)$, where $c$ is the speed of light) for future theoretical modeling.

Detailed information about the sample is provided in Table 1, with the following headings: column (1) source name in the Fermi catalogue; column (2) source name; column (3) redshift; column (4) $\gamma$-ray variability time-scale in units of day; column (5) logarithm of the black hole mass in units of $M_{\odot}$ this symbol is the solar mass; column (6) Bulk Lorentz factor of the jet; column (7) logarithm of the broad-line luminosity in units of ergs$^{-1}$; column (8) observation time from infrared to X-ray bands; (9) integration time of $\gamma$-ray band.

## 3 SED FITTING RESULTS

In empirical fittings of blazar SEDs, both quadratic and cubic functions have been widely employed (Abdo et al. 2010a; Chen 2014; Fan et al. 2016b; Yang et al. 2022a, 2023). Recently, Deng et al. (2024) show that cubic functions work better than quadratic functions for fitting blazar SEDs, whereas for symmetric SEDs the two functions yield comparable results. Statistical comparison shows that cubic functions achieve lower AICc values in most cases (33/45 for the low-energy peak and 35/45 for the high-energy peak), indicating a superior goodness of fit. The extra degree of freedom in cubic

[3] https://fermi.gsfc.nasa.gov/cgi-bin/ssc/LAT/LATDataQuery.cgi
[4] https://pypi.org/project/easyFermi/
[5] https://fermipy.readthedocs.io/en/latest/
[6] https://docs.astropy.org/en/stable/api/astropy.stats.bayesian_blocks.html#bayesian-blocks

**Table 1.** Details of the quasi-simultaneous data of the sample.

| Fermi-LAT name | Source name | $z$ | $\tau_{\rm var}$ | log $M$ | $\Gamma$ | log $L_{\rm BLR}$ | $t_1$ | $t_2$ |
|---|---|---|---|---|---|---|---|---|
| (1) | (2) | (3) | (4) | (5) | (6) | (7) | (8) | (9) |
| 4FGL J0022.5+0608 | PKS 0019+058 | 2.860 | - | - | 14.30$^{\rm C}$ | - | 2009.05.17-2009.05.22 | 2009.04.16-2009.06.16 |
| 4FGL J0050.7-0929 | PKS 0048-09 | 0.635 | - | - | 20.23$^{\rm L}$ | - | 2009.12.23 | 2009.11.23-2010.01.23 |
| 4FGL J0051.1-0648 | PKS 0048-071 | 1.975 | - | $9.31 \pm 0.17^{\rm P}$ | 5.61$^{\rm L}$ | - | 2010.02.05 | 2009.11.20-2010.04.20 |
| 4FGL J0109.7+6133 | TXS 0106+612 | 0.785 | - | $7.48 \pm 0.25^{\rm P}$ | 23.74$^{\rm L}$ | - | 2010.01.31-2010.02.03 | 2010.01.01-2010.03.01 |
| 4FGL J0113.4+4948 | S4 0110+49 | 0.389 | - | $8.23 \pm 0.10^{\rm P}$ | 5.66$^{\rm L}$ | - | 2010.08.27 | 2010.07.27-2010.09.27 |

**Note.**
[P]The black hole mass from Paliya et al. (2021).
[L]The Doppler factor from Liodakis et al. (2018).
[C]The Doppler factor from Chen (2018), see their work for details.
This table provides basic information about the samples, and the derived fitting parameters are given in Table 2.
(The complete table is available in machine-readable form.)

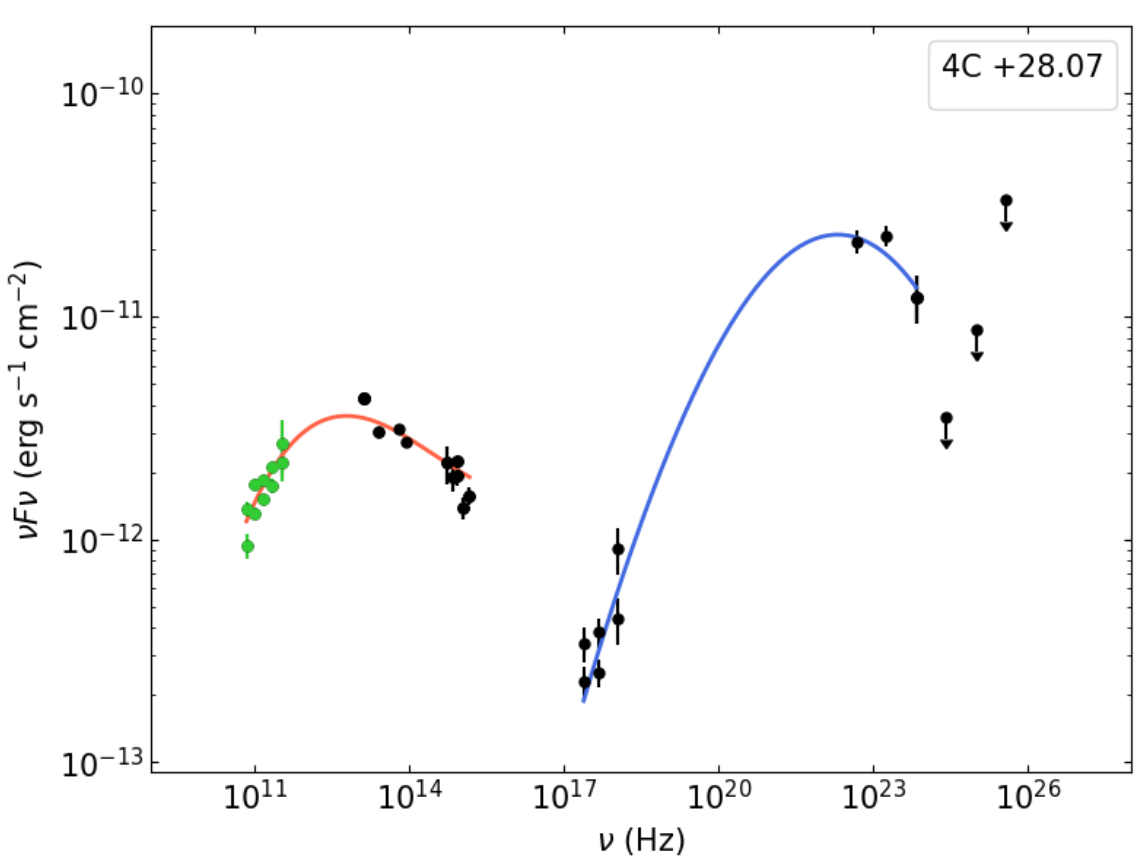

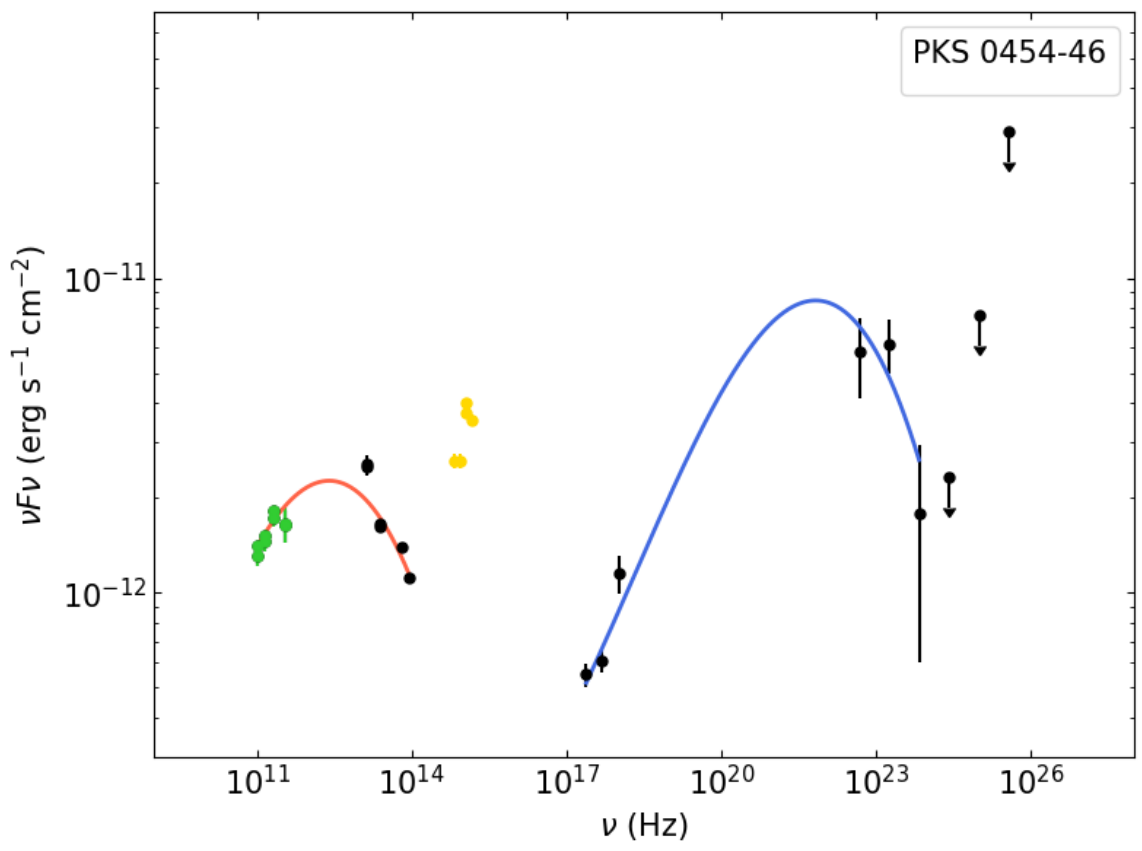

**Figure 2.** The SEDs with fitted lines. The green points represent higher-frequency radio data, the yellow points represent thermal components (either the blue bump or a contribution from the host galaxy), while the black points represent the quasi-simultaneous data, and those with arrows denote upper limits. The red line represents the fit to the synchrotron peak, and the blue line represents the fit to the IC peak.
(The complete figure set (93 images) is available.)

functions allows a more realistic description of asymmetric spectral peaks, in contrast to quadratic functions whose symmetry assumption can cause systematic deviations. Following Deng et al. (2024), we employ cubic function

$$\log\,(\nu F_\nu) = a(\log \nu)^3 + b(\log \nu)^2 + c(\log \nu) + d, \tag{1}$$

where $a$, $b$, $c$ and $d$ are constants, to fit the synchrotron component and IC component, respectively. In the fitting process, the code iminuit is used to obtain the best-fit parameter values and corresponding uncertainties (Dembinski et al. 2025). We also make sure that the fitted curve does not exceed the upper limits, except in cases where the flux of a valid data point adjacent to the upper limit is significantly higher. Then we can calculate the peak frequency ($\log \nu_{\rm syn}^{\rm peak}$, $\log \nu_{\rm IC}^{\rm peak}$), the peak flux ($\nu F_{\nu_{\rm syn}}^{\rm peak}$, $\nu F_{\nu_{\rm IC}}^{\rm peak}$), and the curvature ($\kappa_{\rm syn}^{\rm peak}$, $\kappa_{\rm IC}^{\rm peak}$) at the peak,

$$\kappa_{\rm syn/IC}^{\rm peak} = \left|6a(\log \nu_{\rm syn/IC}^{\rm peak}) + 2b\right|. \tag{2}$$

During the fitting procedure, several considerations should be taken into account. Most importantly, some low synchrotron peak sources (particularly FSRQs) exhibit a prominent blue bump (i.e., thermal emission originating from the accretion disk) in the ultraviolet band. Besides, for some sources such as Mkn 421, there is a contribution from the host galaxy in the optical band. Here, we visually assess whether the thermal emission originates from the accretion disk or the host galaxy, based on whether its peak frequency lies in the optical or ultraviolet band. However, this approach has certain limitations. For example, accretion disk emission might be shifted into the optical band for high-redshift blazars. Nevertheless, what can be confirmed is that these thermal components clearly cannot be attributed to the jet emission and must be excluded when fitting the non-thermal components originated from the relativistic jet. In the SED plot, we mark the thermal component data points in yellow to distinguish them from the non-thermal emission. For ISPs, the X-ray component could originate from either synchrotron radiation or IC scattering, since it lies in the valley between the two humps of the SEDs. In the fitting of the SEDs, the spectral index $\alpha$ ($\nu F_\nu \propto \nu^{-\alpha}$) of X-ray data points can help determine the radiation mechanism of the X-ray component. If $\alpha > 0$, it indicates that the

**Table 2.** SED fitting parameters of synchrotron and IC components.

| **Fermi-LAT name** | **source name** | $a_1$ | $b_1$ | $c_1$ | $d_1$ | $\log\nu_{\rm syn}^{\rm peak}$ | $\kappa_{\rm syn}^{\rm peak}$ | $\log L_{\rm syn}^{\rm peak}$ | $\log L_{\rm bol}$ |
|---|---|---|---|---|---|---|---|---|---|
| (1) | (2) | (3) | (4) | (5) | (6) | (7) | (8) | (9) | (10) |
| 4FGL J0022.5+0608 | PKS 0019+058 | -0.012 ± 4.34e-5 | 0.21 ± 8.4e-4 | 0.61 ± 0.013 | -28.72 ± 0.11 | 13.06 ± 0.066 | 0.51 ± 0.0069 | 47.3 ± 0.26 | 48.60 ± 0.73 |
| 4FGL J0050.7-0929 | PKS 0048-09 | 0.014 ± 7.81e-6 | -0.66 ± 1.59e-4 | 10.3 ± 0.002 | -64.49 ± 0.027 | 13.96 ± 0.04 | 0.16 ± 0.0035 | 46.02 ± 0.058 | 47.18 ± 0.11 |
| 4FGL J0051.1-0648 | PKS 0048-071 | -0.006 ± 6.61e-5 | 0.097 ± 0.0012 | 0.35 ± 0.013 | -19.32 ± 0.12 | 12.01 ± 0.17 | 0.25 ± 0.01 | 46.63 ± 0.28 | 47.89 ± 0.48 |
| 4FGL J0109.7+6133 | TXS 0106+612 | -0.02 ± 2.41e-5 | 0.32 ± 3.91e-4 | 1.07 ± 5.19e-3 | -37.12 ± 0.051 | 12.58 ± 0.02 | 0.81 ± 0.0034 | 46.23 ± 0.11 | 48.16 ± 0.42 |
| 4FGL J0113.4+4948 | S4 0110+49 | -0.009 ± 1.86e-5 | 0.15 ± 3.39e-4 | 0.55 ± 0.005 | -25.14 ± 0.05 | 13.28 ± 0.04 | 0.39 ± 0.003 | 45.40 ± 0.01 | 46.50 ± 0.42 |

| **Fermi-LAT name** | **source name** | $a_2$ | $b_2$ | $c_2$ | $d_2$ | $\log\nu_{\rm IC}^{\rm peak}$ | $\kappa_{\rm IC}^{\rm peak}$ | $\log L_{\rm IC}^{\rm peak}$ | $\chi^2$ |
|---|---|---|---|---|---|---|---|---|---|
| (1) | (2) | (3) | (4) | (5) | (6) | (7) | (8) | (9) | (10) |
| 4FGL J0022.5+0608 | PKS 0019+058 | -0.0057 ± 3.85e-5 | 0.26 ± 0.0012 | -3.33 ± 0.021 | -4.68 ± 0.29 | 21.96 ± 0.35 | 0.22 ± 0.014 | 47.65 ± 0.68 | 4.06 |
| 4FGL J0050.7-0929 | PKS 0048-09* | 0.026 ± 1.27e-5 | -0.94 ± 1.74e-4 | 10.96 ± 0.017 | -52.96 ± 0.17 | 20.3 ± 0.12 | 0.25 ± 0.019 | 46.32 ± 0.49 | 3.23 |
| 4FGL J0051.1-0648 | PKS 0048-071 | -0.002 ± 2.96e-5 | 0.061 ± 8.3e-4 | 0.35 ± 0.016 | -21.62 ± 0.37 | 21.62 ± 0.37 | 0.15 ± 0.0073 | 47.04 ± 0.64 | 1.74 |
| 4FGL J0109.7+6133 | TXS 0106+612 | -0.004 ± 1.72e-5 | 0.14 ± 5.17e-4 | -0.07 ± 0.012 | -32.60 ± 0.16 | 22.01 ± 0.13 | 0.28 ± 0.0047 | 47.31 ± 0.44 | 2.94 |
| 4FGL J0113.4+4948 | S4 0110+49 | -0.0017 ± 3.20e-4 | 0.24 ± 0.015 | -0.049 ± 8.4e-4 | -22.51 ± 0.21 | 21.80± 0.51 | 0.12 ± 0.0078 | 45.39 ± 0.25 | 3.75 |

**Note.**
*The source of unphysical curvature resulting from sparse data points.
This table presents the SED parameters and their associated errors for the synchrotron component (upper table) and the IC component (lower table) within the sample. For both the upper and lower tables, columns (1) and (2) display the source name in the Fermi catalogue and the source name, respectively; columns (3) to (6) display the cubic function coefficients; and columns (7) to (9) display the peak frequency, curvature, and luminosity, respectively. Additionally, column (10) provides the full wavelength bolometric luminosity and the chi-square value for the upper and lower tables, respectively. The chi-square value is calculated using the formula $\chi^2 = \frac{1}{m-{\rm dof}}\sum_{i=1}^{m}\left(\frac{\hat{y}_i - y_i}{\sigma_i}\right)^2$, where $m$ is the number of quasi-simultaneous observational data, dof are the degrees of freedom, $\hat{y}_i$ are the expected values from the model, $y_i$ are the observed data, and $\sigma_i$ is the standard deviation for each data point. In our sample, these errors of data points from infrared to X-ray bands are collected from the SSDC website. For errors of data points that cannot be found, we take 5% of the observed UV and X-ray fluxes as the errors of these data points. The errors of the $\gamma$-ray data points are obtained by the Fermi-LAT data analysis. The luminosity is calculated using the formula $L = 4\pi d_{\rm L}^2 \nu F_\nu$, where $d_{\rm L} = (1+z)\frac{c}{H_0}\int_0^z \frac{1}{\sqrt{\Omega_{\rm M}(1+z)^3+\Omega_\Lambda}}\,{\rm d}z$ is the luminosity distance. The peak luminosity is derived by substituting the peak value of $\nu F_\nu$, while the bolometric luminosity is obtained by integrating the SEDs over frequency.
(The complete table is available in machine-readable form.)

X-ray emission is dominated by synchrotron radiation. Conversely, if $\alpha \le 0$, the X-ray emission is more likely dominated by IC scattering. When the X-ray data points are sparse, it becomes difficult to clearly determine whether the spectrum is rising or falling (e.g., 1RXS J154256.6+6 and 3C 371). In such cases, we typically select the middle portion of these data points, treating the front part as the tail of the synchrotron component and the latter part as the head of the IC component to constrain both. The fit doesn't change much, unless we try to force the IC peak to start much earlier.

The SED fitting results are presented in Figure 2, and the derived parameters, including peak frequency, spectral curvature, peak luminosity, and bolometric luminosity are listed in Table 2.

## 4 CORRELATIONS RESULTS AND DISCUSSIONS

### 4.1 The Distributions

The distributions of derived physical parameters for FSRQs, BL Lacs and BCUs are given in Figure 3. There are only two BCUs, and their parameter values can be directly obtained from Figure 3; the parameter distribution results for FSRQs and BL Lacs are summarized as follows.

We fit the distributions of each parameter using Gaussian functions. The analysis reveals that regarding the peak frequency distribution, the synchrotron peak frequency (mean $\log[\nu_{\rm syn}^{\rm peak}/{\rm Hz}] = 12.79$) and the IC peak frequency (mean $\log[\nu_{\rm IC}^{\rm peak}/{\rm Hz}] = 22.34$) of FSRQs are both centered at relatively lower values than those for BL Lacs (mean $\log[\nu_{\rm syn}^{\rm peak}/{\rm Hz}] = 13.75$, $\log[\nu_{\rm IC}^{\rm peak}/{\rm Hz}] = 22.77$). The synchrotron peak frequency distribution is more concentrated, while the IC peak frequency exhibits a broader range of variation. The curvature distributions are relatively concentrated. The IC peak curvature distribution is centered lower than the synchrotron peak curvature distribution for both types (mean $\kappa^{\rm peak}_{\rm IC} = 0.23$ vs. $\kappa^{\rm peak}_{\rm syn} = 0.39$ for FSRQs, and mean $\kappa^{\rm peak}_{\rm IC} = 0.15$ vs. $\kappa^{\rm peak}_{\rm syn} = 0.40$ for BL Lacs). The overall curvature distribution is left-skewed, with only a very small number of samples exhibiting curvature greater than 0.8. Regarding peak luminosity distributions, both the synchrotron and IC distributions for FSRQs are shifted toward significantly higher values (mean $\log[L^{\rm peak}_{\rm syn}/{\rm erg\ s^{-1}}] = 46.42$, $\log[L^{\rm peak}_{\rm IC}/{\rm erg\ s^{-1}}] = 46.93$) compared to BL Lacs (mean $\log[L^{\rm peak}_{\rm syn}/{\rm erg\ s^{-1}}] = 45.62$, $\log[L^{\rm peak}_{\rm IC}/{\rm erg\ s^{-1}}] = 45.51$). The FSRQs peak luminosity distributions are also narrower. Similarly, the distribution of the bolometric luminosity for FSRQs is shifted higher, and is narrower than that for BL Lacs (mean $\log[L_{\rm bol}/{\rm erg\ s^{-1}}] = 47.90$ for FSRQs and $\log[L_{\rm bol}/{\rm erg\ s^{-1}}] = 46.71$ for BL Lacs). The variability time-scales for both FSRQs and BL Lacs are on the order of days. The mean value of variability time-scale for BL Lacs (mean $\tau_{\rm var} = 6.22$ day) is slightly larger than that for FSRQs (mean $\tau_{\rm var} = 4.17$ day).

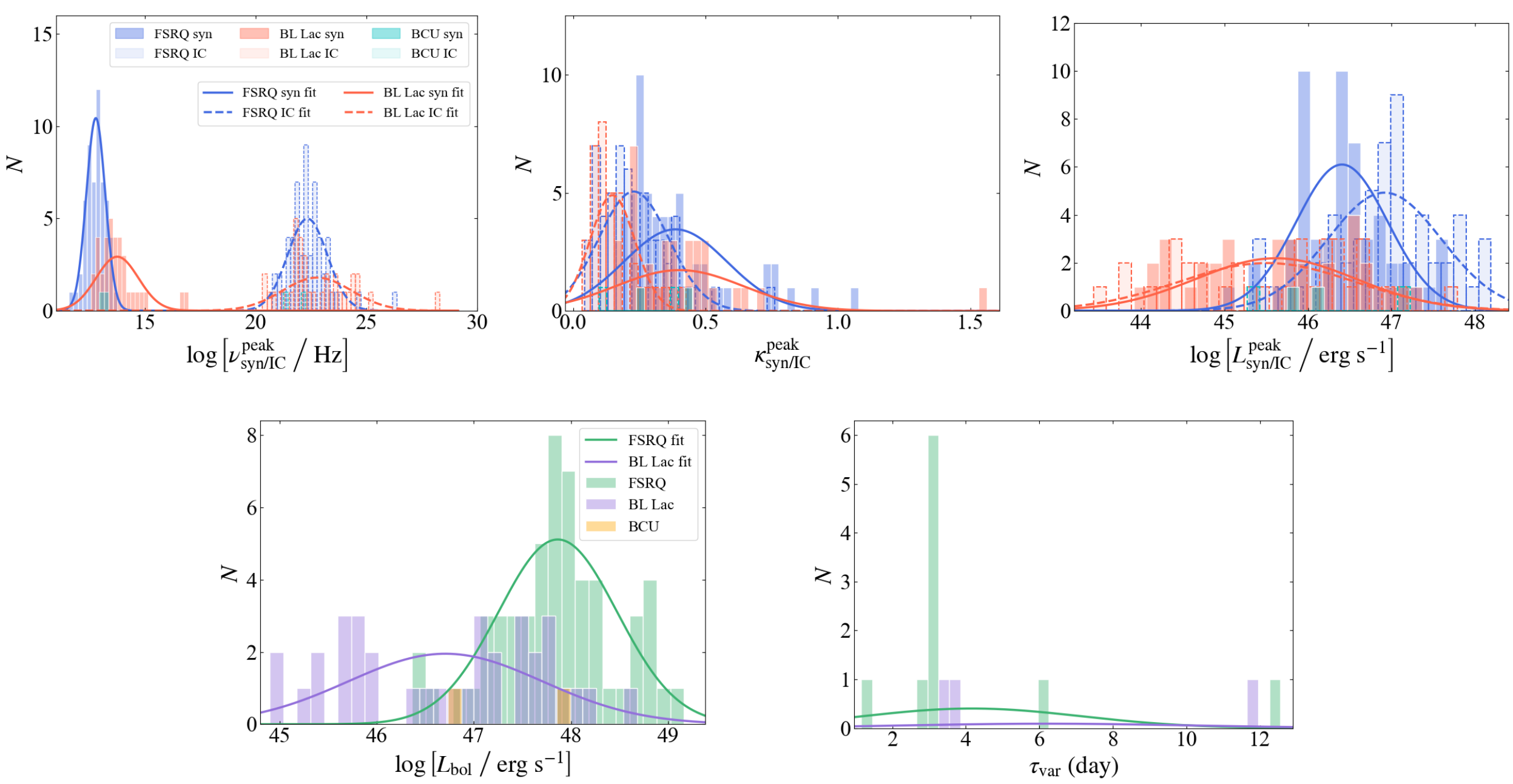

**Figure 3.** Distributions of $\log\nu^{\rm peak}_{\rm syn/IC}$, $\kappa^{\rm peak}_{\rm syn/IC}$, $\log L^{\rm peak}_{\rm syn/IC}$, $\log L_{\rm bol}$, $\tau_{\rm var}$ of FSRQs, BL Lacs and BCUs. For FSRQs, the synchrotron peak frequency ($\log[\nu^{\rm peak}_{\rm syn}/{\rm Hz}]$) lies in the range 11.58 ~ 13.66, with peak curvature ($\kappa^{\rm peak}_{\rm syn}$) in the range 0.18 ~ 1.07. The synchrotron peak luminosity ($\log[L^{\rm peak}_{\rm syn}/{\rm erg\ s^{-1}}]$) is found within 44.98 ~ 47.65. The IC peak frequency ($\log[\nu^{\rm peak}_{\rm IC}/{\rm Hz}]$) lies in the range 20.70 ~ 26.22, with peak curvature ($\kappa^{\rm peak}_{\rm IC}$) in the range 0.04 ~ 0.76. The corresponding IC peak luminosity ($\log[L^{\rm peak}_{\rm IC}/{\rm erg\ s^{-1}}]$) is found within 45.34 ~ 48.16. And the integrated luminosity ($\log[L_{\rm bol}/{\rm erg\ s^{-1}}]$) lies in the range 46.36 ~ 49.04. The variability time-scale ($\tau_{\rm var}$) lies in the range 1.01 ~ 10.20 day. For BL Lacs, the synchrotron peak frequency ($\log[\nu^{\rm peak}_{\rm syn}/{\rm Hz}]$) lies in the range 12.55 ~ 16.84, with peak curvature ($\kappa^{\rm peak}_{\rm syn}$) in the range 0.15 ~ 1.55. The synchrotron peak luminosity ($\log[L^{\rm peak}_{\rm syn}/{\rm erg\ s^{-1}}]$) is found within 43.92 ~ 47.30. The IC peak frequency ($\log[\nu^{\rm peak}_{\rm IC}/{\rm Hz}]$) lies in the range 20.30 ~ 28.12, with peak curvature ($\kappa^{\rm peak}_{\rm IC}$) in the range 0.03 ~ 0.41. The corresponding IC peak luminosity ($\log[L^{\rm peak}_{\rm IC}/{\rm erg\ s^{-1}}]$) is found within 43.44 ~ 47.65. And the integrated luminosity ($\log[L_{\rm bol}/{\rm erg\ s^{-1}}]$) lies in the range 44.90 ~ 48.60. The variability time-scale ($\tau_{\rm var}$) lies in the range 1.16 ~ 12.40 day.

## 4.2 Pairwise Correlation between Physical Quantities

Table 3 displays the pairwise correlations between the parameters, including $\log\nu^{\rm peak}_{\rm syn/IC}$, $\kappa^{\rm peak}_{\rm syn/IC}$, $\log L^{\rm peak}_{\rm syn/IC}$, $\log L_{\rm bol}$, $\log M$, $\Gamma$, and $\log L_{\rm BLR}$. Notably, although BCUs are not formally classified as FSRQs or BL Lacs, the two available BCU sources in our sample are included in the correlation analysis for both sub-populations as well as the total sample to maintain completeness. Due to this limited number, their impact on individual sub-population statistics is minimal. The correlations are categorized into three groups: significant correlations ($p < 0.05$, $r > 0.5$), weak correlations ($p < 0.05$, $r < 0.5$), and non-correlations ($p > 0.05$). For significant correlations, the best linear equations are summarized in Table 4, together with their corresponding $r$ and $p$ values.

Among the above significant correlations, some are also found in previous studies with different samples. For the correlation between $\log L^{\rm peak}_{\rm syn}$ and $\log L^{\rm peak}_{\rm IC}$ (see the top panel of Fig. 4), our best linear fitting results are

$$\begin{aligned}
\text{FSRQs:}\quad & \log L^{\rm peak}_{\rm IC} = (1.12\pm0.11)\log L^{\rm peak}_{\rm syn} + (-0.52\pm5.16),\\
\text{BL Lacs:}\quad & \log L^{\rm peak}_{\rm IC} = (1.04\pm0.12)\log L^{\rm peak}_{\rm syn} + (-1.85\pm2.91),\\
\text{Whole:}\quad & \log L^{\rm peak}_{\rm IC} = (1.22\pm0.07)\log L^{\rm peak}_{\rm syn} + (-10.02\pm3.33).
\end{aligned} \tag{3}$$

Similar correlations have also been reported in the literature. For

**Table 3.** Correlation between physical quantities

| | $\log \nu_{\rm syn}^{\rm peak}$ | $\log \nu_{\rm IC}^{\rm peak}$ | $\kappa_{\rm syn}^{\rm peak}$ | $\kappa_{\rm IC}^{\rm peak}$ | $\log L_{\rm syn}^{\rm peak}$ | $\log L_{\rm IC}^{\rm peak}$ | $\log L_{\rm bol}$ | $\log M$ | $\Gamma$ | $\log L_{\rm BLR}$ | $\tau_{\rm var}$ |
|---|---|---|---|---|---|---|---|---|---|---|---|
| $\log \nu_{\rm syn}^{\rm peak}$ | \ | | | | | | | | | | |
| $\log \nu_{\rm IC}^{\rm peak}$ | ✓/×/△ | \ | | | | | | | | | |
| $\kappa_{\rm syn}^{\rm peak}$ | ×/×/× | △/×/× | \ | | | | | | | | |
| $\kappa_{\rm IC}^{\rm peak}$ | ×/×/× | ×/✓/× | ×/×/× | \ | | | | | | | |
| $\log L_{\rm syn}^{\rm peak}$ | ×/✓/✓ | ×/×/× | △/×/× | △/×/△ | \ | | | | | | |
| $\log L_{\rm IC}^{\rm peak}$ | ×/△/△ | ×/×/× | ×/×/× | ✓/×/✓ | ✓/✓/✓ | \ | | | | | |
| $\log L_{\rm bol}$ | ×/△/✓ | ×/×/× | △/×/× | △/×/△ | ✓/✓/✓ | ✓/✓/✓ | \ | | | | |
| $\log M$ | ✓/×/✓ | ×/×/× | ×/×/× | ×/×/× | ✓/×/✓ | ×/×/× | ×/×/× | \ | | | |
| $\Gamma$ | ×/×/× | ×/×/× | ×/×/× | ×/×/× | ×/×/× | ×/×/× | ×/×/× | ×/×/× | \ | | |
| $\log L_{\rm BLR}$ | ×/×/× | ✓/×/△ | ×/×/× | ×/×/× | ×/×/× | ×/×/× | ×/×/× | ✓/×/✓ | ×/×/✓ | \ | |
| $\tau_{\rm var}$ | ×/×/× | ×/×/× | ×/×/× | ×/×/× | ×/×/✓ | ×/×/✓ | ×/×/✓ | ×/×/× | ×/×/× | ×/×/× | \ |

**Note.** For each cell, the result is split into three parts separated by slashes: the left part indicates the statistical significance for FSRQs, the middle part indicates that for BL Lacs, and the right part indicates that for the whole sample. A tick (✓) denotes a relatively significant correlation ($|r| > 0.5$, $p < 0.05$), a triangle (△) denotes a weak correlation ($|r| < 0.5$, $p < 0.05$), and a cross (×) denotes no significant correlation ($p > 0.05$).

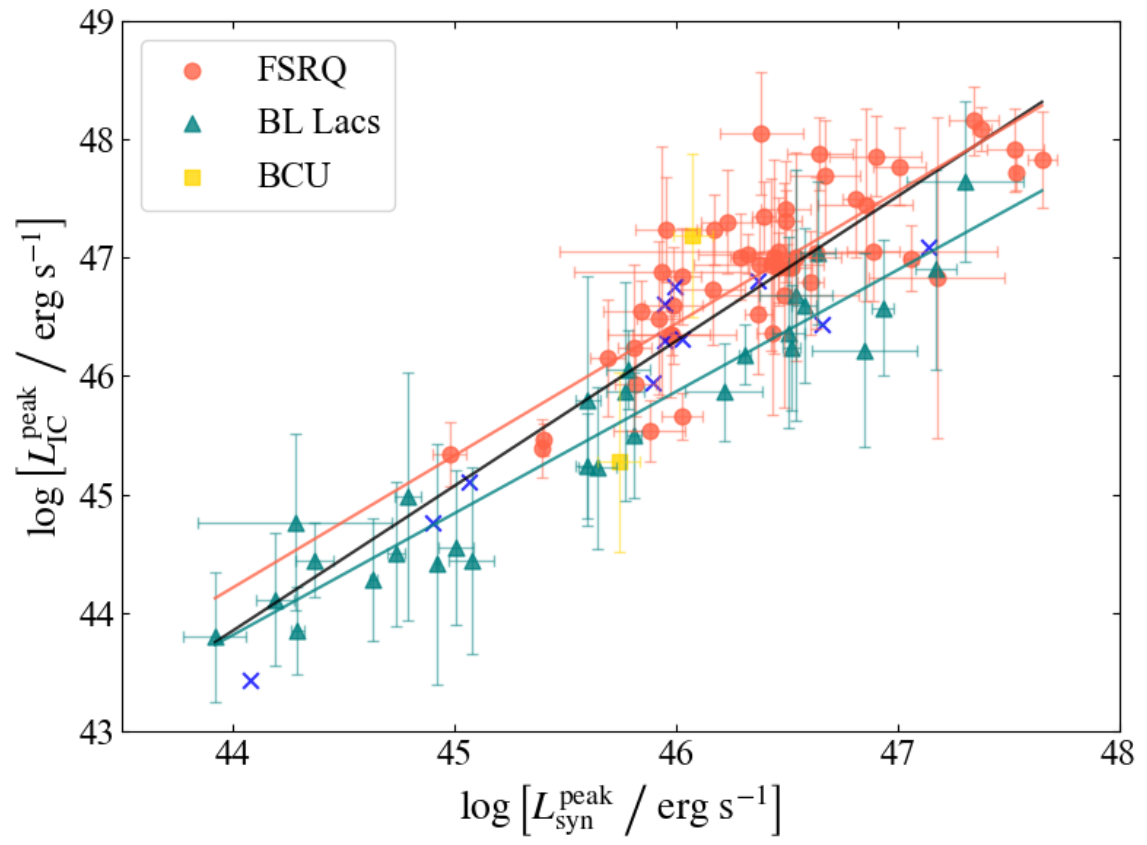

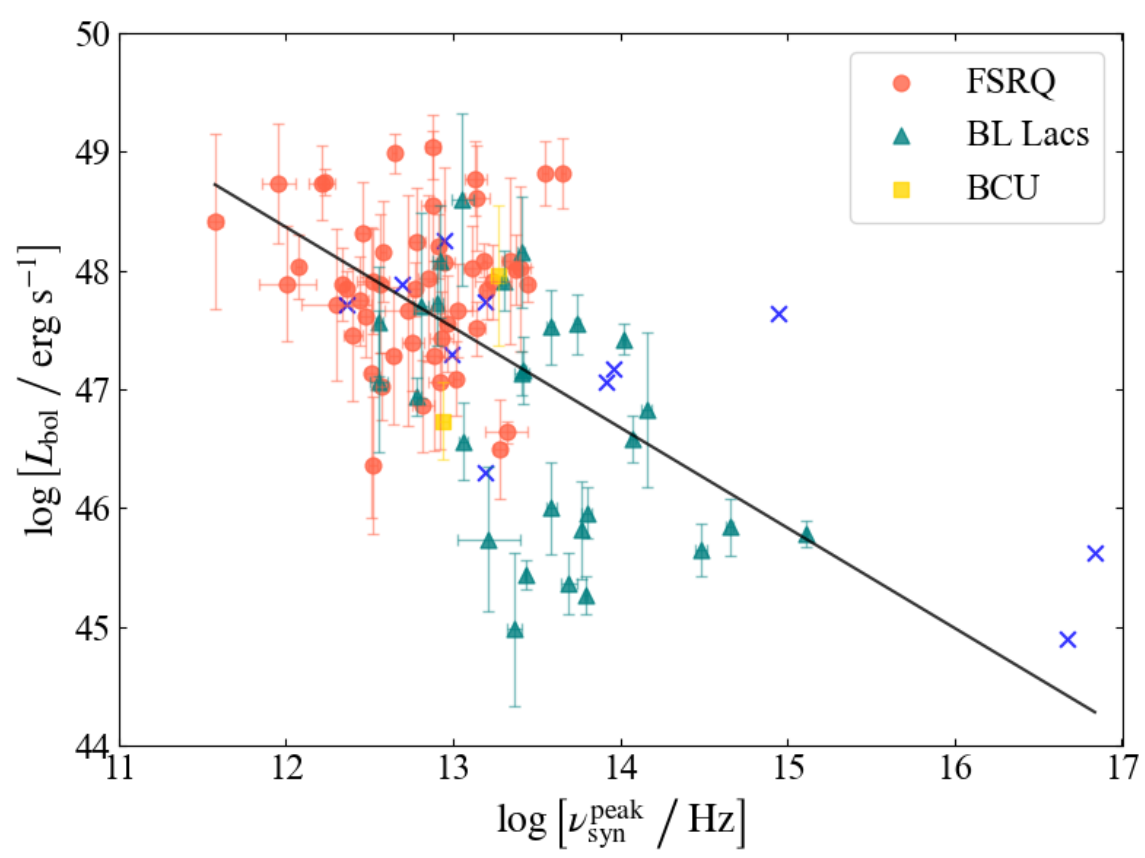

**Figure 4.** Top panel: correlation between the synchrotron peak luminosity and the IC peak luminosity. Bottom panel: correlation between the synchrotron peak frequency and the bolometric luminosity. The blue crosses indicate the source of unphysical curvature resulting from sparse data points (see Table 2). The black, red, and green solid lines are the best linear fitting for the full sample, FSRQs, and BL Lacs, respectively. Please note that the best linear fitting equation is only shown when a correlation is found ($p < 0.05$).

instance, Xue et al. (2016) reported

$$\begin{aligned} \text{FSRQs:} \quad & \log L_{\rm IC}^{\rm peak} = (1.45 \pm 0.11) \log L_{\rm syn}^{\rm peak} + (-20.10 \pm 5.12), \\ \text{BL Lacs:} \quad & \log L_{\rm IC}^{\rm peak} = (1.12 \pm 0.10) \log L_{\rm syn}^{\rm peak} + (-5.43 \pm 4.61), \end{aligned} \tag{4}$$

based on quasi-simultaneous data; Yang et al. (2023) reported

$$\begin{aligned} \text{FSRQs:} \quad & \log L_{\rm IC}^{\rm peak} = (1.10 \pm 0.03) \log L_{\rm syn}^{\rm peak} + (-4.23 \pm 1.32), \\ \text{BL Lacs:} \quad & \log L_{\rm IC}^{\rm peak} = (1.01 \pm 0.02) \log L_{\rm syn}^{\rm peak} + (-0.63 \pm 0.74), \\ \text{Whole:} \quad & \log L_{\rm IC}^{\rm peak} = (1.14 \pm 0.02) \log L_{\rm syn}^{\rm peak} + (-6.28 \pm 0.70), \end{aligned} \tag{5}$$

and Zhang et al. (2024a) reported

$$\log L_{\rm IC}^{\rm peak} = (1.35 \pm 0.03) \log L_{\rm syn}^{\rm peak} + (-15.40 \pm 1.56), \tag{6}$$

for FSRQs. The slopes of the best fitting equations obtained in this work are generally consistent with the above previous studies. Furthermore, for the correlation between $\log L_{\rm bol}$ and $\log \nu_{\rm syn}^{\rm peak}$ (see the bottom panel of Fig. 4), our fitting result for the whole sample is

$$\log L_{\rm bol} = (-0.82 \pm 0.15) \log \nu_{\rm syn}^{\rm peak} + (58.60 \pm 2.09). \tag{7}$$

The negative correlation found in our work may support the phenomena of the blazar sequence (Fossati et al. 1998). In contrast, Fan et al. (2016b) reported

$$\log L_{\rm bol} = (0.06 \pm 0.03) \log \nu_{\rm syn}^{\rm peak} + (44.58 \pm 0.41), \tag{8}$$

which is markedly different from our result. The difference may arise primarily from the use of non-simultaneous data in their sample compared to the use of quasi-simultaneous data in our study. Although our working sample is simultaneous, it is highly biased because of the relatively small sample size and the dominance of FSRQs. However, the significant correlations revealed by the sample may still suggest genuine physical correlations between these quantities. Therefore, for those blazars whose SEDs lack data in a specific waveband, the best linear fitting equations described above might be utilized to estimate the values of the corresponding physical quantities.

Regardless of whether these correlations are statistically significant, many of them may carry important physical implications or

**Table 4.** Correlation analysis results.

| Correlation (x vs. y)<br>(1) | Sample<br>(2) | $\beta$<br>(3) | $\alpha$<br>(4) | $r$<br>(5) | $p$<br>(6) |
|---|---|---|---|---|---|
| $\log \nu_{\rm syn}^{\rm peak}$ vs. $\log L_{\rm syn}^{\rm peak}$ | Whole | $-0.55 \pm 0.15$ | $53.33 \pm 1.90$ | -0.52 | $< 0.0001$ |
| $\log \nu_{\rm syn}^{\rm peak}$ vs. $\log L_{\rm bol}$ | Whole | $-0.82 \pm 0.15$ | $58.60 \pm 2.09$ | -0.49 | $< 0.0001$ |
| $\log \nu_{\rm syn}^{\rm peak}$ vs. $\log M$ | Whole | $-0.78 \pm 0.40$ | $18.53 \pm 5.11$ | -0.53 | 0.005 |
| $\kappa_{\rm IC}^{\rm peak}$ vs. $\log L_{\rm IC}^{\rm peak}$ | Whole | $4.71 \pm 0.89$ | $45.50 \pm 0.22$ | 0.65 | $< 0.0001$ |
| $\log L_{\rm syn}^{\rm peak}$ vs. $\log L_{\rm IC}^{\rm peak}$ | Whole | $1.22 \pm 0.07$ | $-10.02 \pm 3.33$ | 0.86 | $< 0.0001$ |
| $\log L_{\rm syn}^{\rm peak}$ vs. $\log L_{\rm bol}$ | Whole | $1.13 \pm 0.04$ | $-4.57 \pm 1.86$ | 0.92 | $< 0.0001$ |
| $\log L_{\rm syn}^{\rm peak}$ vs. $\log M$ | Whole | $0.57 \pm 0.11$ | $-17.97 \pm 5.26$ | 0.61 | $< 0.0001$ |
| $\log L_{\rm IC}^{\rm peak}$ vs. $\log L_{\rm bol}$ | Whole | $0.89 \pm 0.05$ | $6.05 \pm 2.23$ | 0.98 | $< 0.0001$ |
| $\log M$ vs. $\log L_{\rm BLR}$ | Whole | $1.10 \pm 0.20$ | $35.07 \pm 1.72$ | 0.63 | 0.001 |
| $\Gamma$ vs. $\log L_{\rm BLR}$ | Whole | $0.70 \pm 0.23$ | $38.71 \pm 2.03$ | 0.50 | 0.02 |
| $\log L_{\rm syn}^{\rm peak}$ vs. $\tau_{\rm var}$ | Whole | $-2.82 \pm 0.96$ | $135.77 \pm 44.40$ | -0.66 | 0.014 |
| $\log L_{\rm IC}^{\rm peak}$ vs. $\tau_{\rm var}$ | Whole | $-2.49 \pm 0.76$ | $121.33 \pm 35.57$ | -0.57 | 0.04 |
| $\log L_{\rm bol}$ vs. $\tau_{\rm var}$ | Whole | $-2.50 \pm 0.86$ | $124.45 \pm 41.36$ | -0.59 | 0.036 |
| $\log \nu_{\rm syn}^{\rm peak}$ vs. $\log \nu_{\rm IC}^{\rm peak}$ | FSRQ | $0.61 \pm 0.19$ | $14.24 \pm 2.42$ | 0.58 | $< 0.0001$ |
| $\log \nu_{\rm syn}^{\rm peak}$ vs. $\log M$ | FSRQ | $-1.31 \pm 0.59$ | $25.13 \pm 7.50$ | -0.60 | 0.008 |
| $\log \nu_{\rm IC}^{\rm peak}$ vs. $\log L_{\rm BLR}$ | FSRQ | $-0.54 \pm 0.30$ | $56.59 \pm 6.60$ | -0.53 | 0.02 |
| $\kappa_{\rm IC}^{\rm peak}$ vs. $\log L_{\rm IC}^{\rm peak}$ | FSRQ | $2.90 \pm 0.81$ | $46.24 \pm 0.22$ | 0.57 | $< 0.0001$ |
| $\log L_{\rm syn}^{\rm peak}$ vs. $\log L_{\rm IC}^{\rm peak}$ | FSRQ | $1.12 \pm 0.11$ | $-0.52 \pm 5.16$ | 0.72 | $< 0.0001$ |
| $\log L_{\rm syn}^{\rm peak}$ vs. $\log L_{\rm bol}$ | FSRQ | $0.97 \pm 0.08$ | $2.99 \pm 3.85$ | 0.88 | $< 0.0001$ |
| $\log L_{\rm syn}^{\rm peak}$ vs. $\log M$ | FSRQ | $0.54 \pm 0.13$ | $-16.21 \pm 6.07$ | 0.53 | 0.01 |
| $\log L_{\rm IC}^{\rm peak}$ vs. $\log L_{\rm bol}$ | FSRQ | $0.85 \pm 0.02$ | $7.01 \pm 3.34$ | 0.98 | $< 0.0001$ |
| $\log M$ vs. $\log L_{\rm BLR}$ | FSRQ | $0.70 \pm 0.23$ | $38.71 \pm 2.03$ | 0.67 | 0.02 |
| $\log \nu_{\rm syn}^{\rm peak}$ vs. $\log L_{\rm syn}^{\rm peak}$ | BL Lac | $-0.74 \pm 0.30$ | $55.65 \pm 4.07$ | -0.50 | 0.009 |
| $\log \nu_{\rm IC}^{\rm peak}$ vs. $\kappa_{\rm IC}^{\rm peak}$ | BL Lac | $-0.06 \pm 0.01$ | $1.52 \pm 0.27$ | -0.58 | 0.003 |
| $\log L_{\rm syn}^{\rm peak}$ vs. $\log L_{\rm IC}^{\rm peak}$ | BL Lac | $1.04 \pm 0.12$ | $-1.85 \pm 2.91$ | 0.95 | $< 0.0001$ |
| $\log L_{\rm syn}^{\rm peak}$ vs. $\log L_{\rm bol}$ | BL Lac | $1.03 \pm 0.04$ | $-0.41 \pm 1.78$ | 0.98 | $< 0.0001$ |
| $\log L_{\rm IC}^{\rm peak}$ vs. $\log L_{\rm bol}$ | BL Lac | $0.94 \pm 0.03$ | $4.10 \pm 1.19$ | 0.98 | $< 0.0001$ |

**Note.** Columns from left to right: (1) the pair of parameters for which the correlation is analyzed, presented in the form of the independent variable (x) vs. the dependent variable (y); (2) the sample; (3) slope; (4) intercept; (5) correlation coefficient; (6) chance probability.

have been widely explored in previous studies. We will discuss them further in subsequent subsections.

### 4.3 Correlations between Peak Frequencies and curvature

Through analysis of the correlation between the synchrotron peak frequency ($\log \nu_{\rm sy}$) and curvature ($1/b_{\rm sy}$), Chen (2014) proposed a criterion based on the linear regression slope ($\beta$) to distinguish acceleration mechanisms. Specifically, $\beta \approx 2.5$ indicates statistical acceleration with energy-dependent probability acceleration (EDPA), $\beta \approx 3.33$ indicates statistical acceleration with fluctuation of fractional gain acceleration (FFGA), and $\beta \approx 2.0$ indicates stochastic acceleration. It should be noted that $b_{\rm sy}$ in Chen (2014) is derived from the log-parabolic fit. In this work, we use cubic function to fit the SED, thereby extending the method above to determine the acceleration mechanism by analyzing the peak frequency–curvature correlation using $2/\kappa_{\rm syn/IC}^{\rm peak}$. We employ Spearman test to assess correlations and use Bayesian linear regression to estimate the slope, intercept, and their uncertainties. The correlation between peak frequency and curvature is shown in Figure 5. For the whole sample, no significant correlation is found between synchrotron peak frequency and curvature ($p = 0.23$, $r = -0.25$), as well as between IC peak frequency and curvature ($p = 0.15$, $r = 0.22$). For FSRQs, no significant correlation exists between synchrotron peak frequency and curvature ($p = 0.61$, $r = -0.13$) and no significant correlation is found between IC peak frequency and curvature ($p = 0.36$, $r = 0.17$). For BL Lacs, no significant correlation is found between synchrotron peak frequency and curvature ($p = 0.95$, $r = 0.02$). In contrast, a significant positive correlation is observed between IC peak frequency and curvature ($p = 0.003$, $r = 0.64$), with the best linear fit given by

$$2/\kappa_{\rm IC}^{\rm peak} = (6.04 \pm 1.43) \log \nu_{\rm IC}^{\rm peak} + (-118.61 \pm 31.84). \quad (9)$$

Some previous studies reported a strong positive correlation between synchrotron peak frequency and curvature, though the obtained slope differs. Chen (2014) found significant correlation between syn-chrotron peak frequency and curvature, and obtained $\beta = 2.04\pm0.03$, i.e., stochastic acceleration, in 48 Fermi Bright blazars with simultaneous data, but no significant correlation between IC peak frequency and curvature and ($p = 0.0516$) is found. Xue et al. (2016) collected a quasi-simultaneous sample including 179 FSRQs and 74 BL Lacs from the second LAT AGN catalogue, and found $\beta = 3.69 \pm 0.24$ (close to the FFGA) for FSRQs and $\beta = 1.87 \pm 0.19$, i.e., stochastic acceleration, for BL Lacs. Tan et al. (2020) found $\beta = 3.22 \pm 0.52$, i.e., FFGA, based on a sample of 60 FSRQs with quasi-simultaneous data. Subsequent studies with substantially larger samples, built from archive SEDs, revealed more complex trends, though not always consistent with each other. Yang et al. (2023) used a sample of 2709 blazars from Fermi-LAT 4FGL-DR3 and found $\beta = 1.97 \pm 0.03$, i.e., stochastic acceleration, for all blazars; $\beta = 2.86 \pm 0.04$ (close to the EDPA) for FSRQs; and $\beta = 2.14 \pm 0.11$, i.e., stochastic acceleration, for BL Lacs. Notably, their $\beta$ value for FSRQs was lower than that reported by Xue et al. (2016) and Tan et al. (2020). Chen et al. (2023) analyzed another large sample (including 504 FSRQs, 277 BL Lac objects, and 17 $\gamma$NLS1s) and found $\beta = 2.62 \pm 0.05$ (close to the EDPA) for all blazars, $\beta = 2.63 \pm 0.04$ (close to the EDPA) for FSRQs, and $\beta = 2.55 \pm 0.14$, i.e., EDPA, for BL Lacs. Despite also having a large sample size, their results differed from Yang et al. (2023), they found significantly higher $\beta$ values for the whole sample and BL Lacs, while the $\beta$ of FSRQs was much closer. Xiao et al. (2024b), also based on a large 4FGL-DR3 sample (2705 blazars), found that the overall fit showed $\beta = 2.39\pm0.03$ (close to the EDPA), a value intermediate between the two large sample studies above. Notably, Xiao et al. (2024b) divided the peak frequencies into bins and examined the slope for each bin, finding $\beta$ decreased from $3.72 \pm 0.12$ to $2.56 \pm 0.04$, then increased to $3.80 \pm 0.19$, revealing a transition in acceleration mechanisms from FFGA to EDPA around $\log \nu_{\rm syn}^{\rm peak} \sim 15$. Anjum et al. (2025) found a negative correlation between the synchrotron frequency ($\log \nu_{\rm s}$) and the curvature ($b_{\rm s}$). The slopes differ slightly between BL Lacs ($b_{\rm s} = -0.03 \log \nu_{\rm s} + 0.56$) and FSRQs ($b_{\rm s} = -0.07 \log \nu_{\rm s} + 1.08$). This indicates the stochastic nature of particle acceleration in blazar jets.

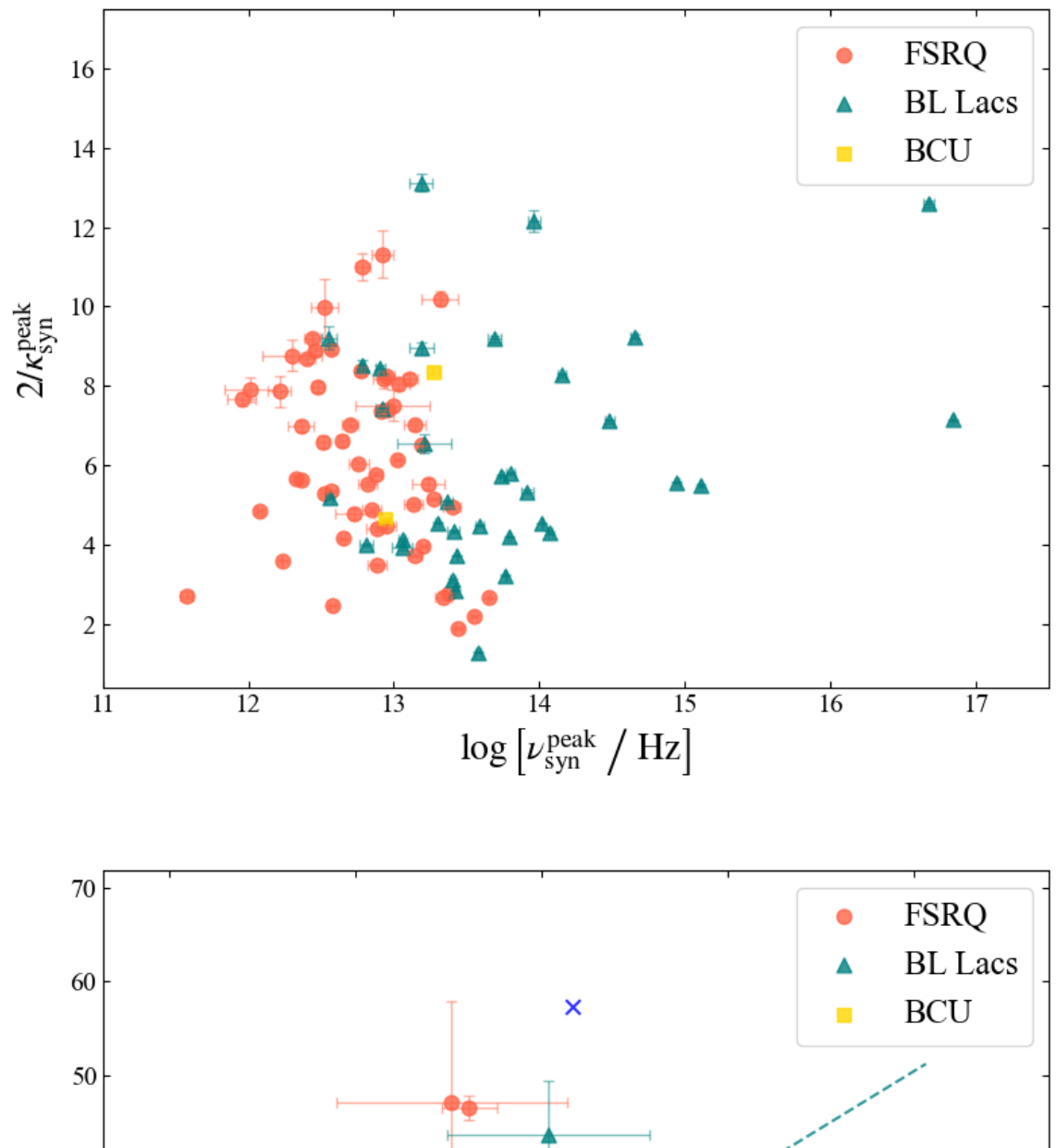

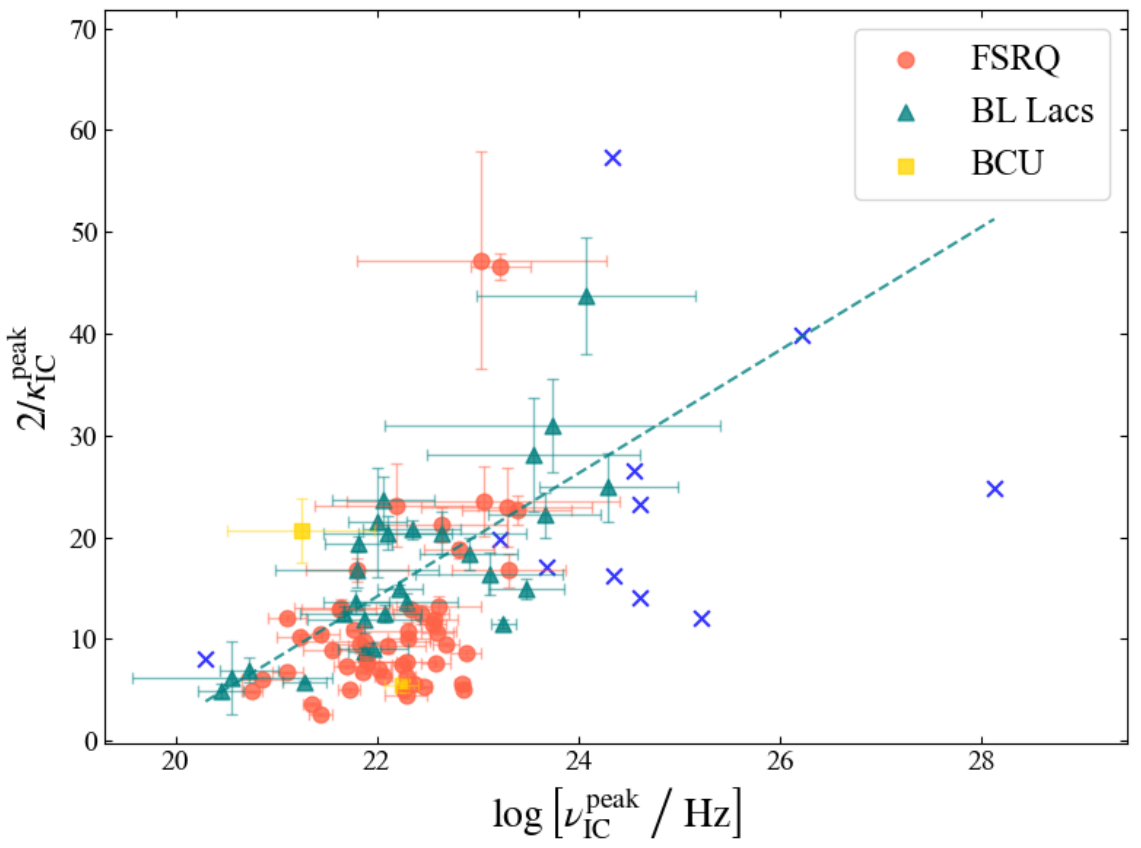

**Figure 5.** Top panel: correlation between the synchrotron peak frequency and the curvature. Bottom panel: correlation between the IC peak frequency and the curvature. The green dashed line is the best linear fitting for the IC peak frequency and curvature of BL Lacs. Please note that the best linear fitting equation is only shown when a correlation is found ($p < 0.05$).

The aforementioned three acceleration mechanisms essentially belong to second-order Fermi acceleration (i.e., stochastic acceleration), differing primarily in their energy-gain processes. Beyond stochastic acceleration, shock acceleration and magnetic reconnection acceleration are prevalent in blazar jets (e.g., Raiteri 2025). The absence of synchrotron peak frequency-curvature correlation in our analysis (after filtering for simultaneous data) suggests stochastic acceleration is not the dominant mechanism. The absence of correlation between the IC peak frequency and curvature for FSRQs may stem from the joint effects of its composite radiation (SSC/EC), and the difference in scattering regime. Furthermore, the limited sample size may also be a contributing factor to the lack of a statistically significant correlation.

### 4.4 Blazar Sequence

The blazar sequence, characterized by the negative correlations between $\log L_{\rm bol}$ and $\log \nu_{\rm syn}^{\rm peak}$, as well as between the Compton dominance $\log Y$ and $\log \nu_{\rm syn}^{\rm peak}$, corresponding to measurements in the observer's frame, has been widely adopted as a phenomenological framework to describe the SEDs of blazars (Fossati et al. 1998; Chen & Bai 2011; Finke 2013; Xiong et al. 2015; Fan et al. 2016b; Ghisellini et al. 2017; Wan et al. 2024). It is interesting that these negative correlations have been shown to be positive when corrected for relativistic Doppler boosting (Nieppola et al. 2008; Wu et al. 2009; Xiong et al. 2015; Fan et al. 2017; Yang et al. 2022b; Ouyang et al. 2023; Wan et al. 2024), corresponding to measurements in the comoving frame. The transformation from the observer's frame to the comoving frame can be approximated as: $\nu \simeq \nu^{\rm obs}\delta^{-1}$, $L \simeq L^{\rm obs}\delta^{-4}$, where the superscript "obs" denotes quantities measured in the observer's frame. Detailed analysis of these correlations offers valuable information on the physical conditions and emission mechanisms within the blazar jet (Ghisellini et al. 1998; Björnsson 2010; Finke 2013; Potter & Cotter 2013; Wan et al. 2024). Given the strong variability of blazar emission, it is essential to examine these correlations using samples constructed from quasi-simultaneous multiwavelength observations to ensure the reliability and physical consistency of the inferred trends. For the needed Doppler factor $\delta$ in our analysis, we first determine $\delta$ for each source, adopting values from Liodakis et al. (2018) when available; otherwise, we use values from Chen (2018). If any of these additional values are extreme ($\delta > 100$ or $\delta < 1$), we replace them with those from Zhang et al. (2020) when possible; if no such value exists, we adopt the average $\delta = 14.3$ reported by Chen (2018). In the following, we revisit the blazar sequence using our sample described in Section 2. All correlation results are listed in Table 5 and Figure 6.

For the correlation between $\log Y$ and $\log \nu_{\rm syn}^{\rm peak}$, we find a moderate

**Table 5.** Correlation results within context of blazar sequence in our sample

| in the observers' frame | N | $\log(Y)$ vs. $\log(\nu_{syn}^{peak})$ | | | $\log(L_{bol})$ vs. $\log(\nu_{syn}^{peak})$ | | |
|---|---|---|---|---|---|---|---|
| | | $r$ | $p$ | $s$ | $r$ | $p$ | $s$ |
| (1) | (2) | (3) | (4) | (5) | (6) | (7) | (8) |
| ALL Blazars | 82 | -0.41 | < 0.0001 | $-0.37 \pm 0.11$ | -0.49 | < 0.0001 | $-0.82 \pm 0.15$ |
| FSRQs | | | | | | | |
| ALL | 52 | -0.15 | 0.31 | $-0.21 \pm 0.18$ | -0.06 | 0.70 | $-0.02 \pm 0.22$ |
| $Y \leq 1$ | 6 | 0.84 | 0.09 | $0.45 \pm 0.98$ | 0.34 | 0.45 | $0.67 \pm 1.52$ |
| $Y > 1$ | 46 | -0.12 | 0.46 | $-0.19 \pm 0.17$ | -0.10 | 0.58 | $-0.02 \pm 0.22$ |
| BL Lacs | | | | | | | |
| ALL | 28 | -0.17 | 0.41 | $-0.03 \pm 0.24$ | -0.47 | 0.03 | $-0.77 \pm 0.30$ |
| $Y \leq 1$ | 18 | 0.03 | 0.92 | $0.07 \pm 0.35$ | -0.28 | 0.30 | $-0.63 \pm 0.56$ |
| $Y > 1$ | 10 | 0.01 | 0.98 | $0.05 \pm 0.40$ | -0.71 | 0.06 | $-0.86 \pm 0.38$ |
| in the comoving frame | | $\log(Y)$ vs. $\log(\nu_{syn}^{peak})$ | | | $\log(L_{bol})$ vs. $\log(\nu_{syn}^{peak})$ | | |
| ALL Blazars | 82 | -0.40 | < 0.0001 | $-0.32 \pm 0.09$ | 0.24 | 0.06 | $0.49 \pm 0.24$ |
| FSRQs | | | | | | | |
| ALL | 52 | -0.23 | 0.13 | $-0.24 \pm 0.13$ | 0.42 | 0.006 | $1.07 \pm 0.31$ |
| $Y \leq 1$ | 6 | 0.51 | 0.32 | $0.27 \pm 0.73$ | 0.69 | 0.12 | $2.19 \pm 2.04$ |
| $Y > 1$ | 46 | -0.15 | 0.36 | $-0.15 \pm 0.14$ | 0.36 | 0.09 | $1.03 \pm 0.33$ |
| BL Lacs | | | | | | | |
| ALL | 28 | -0.26 | 0.19 | $-0.08 \pm 0.17$ | 0.57 | 0.004 | $1.11 \pm 0.37$ |
| $Y \leq 1$ | 18 | -0.03 | 0.90 | $-0.04 \pm 0.23$ | 0.66 | 0.004 | $1.57 \pm 0.56$ |
| $Y > 1$ | 10 | -0.07 | 0.86 | $-0.03 \pm 0.33$ | 0.35 | 0.48 | $0.67 \pm 0.71$ |

**Note.** Columns from left to right: (1): the sample studied in correlation analysis; (2): the number of blazars in the sample; (3) and (6) are the Spearman test correlation coefficients; (4) and (7) are the chance probabilities; (5) and (8) are the slopes of the best linear fitting equations.

negative trend in the observer's frame ($p < 0.0001$, $r = -0.41$, $s = -0.37 \pm 0.11$, where $s$ denotes the slope of the best linear fit) for the entire blazar sample, consistent with previous studies (Fossati et al. 1998; Chen & Bai 2011). A negative correlation is also present in the comoving frame ($p = 0.0001$, $r = -0.40$, $s = -0.32 \pm 0.09$), in agreement with Wan et al. (2024) using the sample with available Doppler factors from Yang et al. (2022b), although the correlation reported there was weaker. When considering FSRQs separately (52 out of 82 blazars), no correlation is found in the observer's frame ($p = 0.31$), consistent with Finke (2013) based on the second *Fermi*-LAT AGN catalogue. No correlation is detected in the comoving frame either ($p = 0.13$). It should be noted that a direct comparison with our previous work (Wan et al. 2024) is difficult because the sample with available Doppler factors in that study was dominated by FSRQs with $Y > 1$ and BL Lacs with $Y > 1$. For BL Lacs (28 out of 82 blazars), no correlation is found in the observer's frame ($p = 0.41$), a result that differs from Finke (2013); Wan et al. (2024), likely due to the smaller sample size in our analysis. Similarly, no significant correlation is found in the comoving frame ($p = 0.19$).

For the correlation between $\log L_{bol}$ and $\log \nu_{syn}^{peak}$, we find a moderate negative trend in the observer's frame ($p < 0.0001$, $r = -0.49$, $s = -0.82 \pm 0.15$) for the entire blazar sample, consistent with numerous previous studies (Fossati et al. 1998; Chen & Bai 2011; Finke 2013; Xiong et al. 2015; Fan et al. 2016b; Ghisellini et al. 2017; Wan et al. 2024). In the comoving frame, no significant correlation is detected (p = 0.06), in contrast to several previous studies (Nieppola et al. 2008; Wu et al. 2009; Fan et al. 2016a; Yang et al. 2022b; Wan et al. 2024). However, the subclass samples (FSRQ and BL Lacs) reveal a significant correlation and similar slopes of the best linear fitting equations, which may be attributed to the scarcity of sources of high peak frequency in the full dataset, thus reducing the overall statistical significance. We emphasize that our primary focus is the correlation between the full-band radiative luminosity and the synchrotron peak frequency, as this choice ensures physical consistency with the theoretical framework proposed by Wan et al. (2024). When considering FSRQs separately, no correlation is found in the observer's frame ($p = 0.70$), whereas the comoving frame shows a moderate positive correlation ($p = 0.006$, $r = 0.42$, $s = 1.07 \pm 0.31$). For BL Lacs, a moderate negative correlation is observed in the observer's frame ($p = 0.03$, $r = -0.47$, $s = -0.77 \pm 0.30$) , while the comoving frame exhibits a moderate positive correlation ($p = 0.004$, $r = 0.57$, $s = 1.11 \pm 0.37$). The subclass-specific correlations in the observer's frame are consistent with results (Ghisellini et al. 2017) obtained using the third *Fermi*-LAT AGN catalogue (3LAC), but differ from those of Wan et al. (2024), who reported correlations for FSRQs rather than BL Lacs.

Based on the theoretical model in our previous work (Wan et al. 2024), FSRQs and BL Lacs with $Y \leq 1$ and $Y > 1$ may exhibit different correlations, driven by distinct physical conditions and emission mechanisms within their jets and radiation regions. For instance, the presence or absence of a correlation, along with the slope of the best linear fit, can provide insights into whether the jet is accelerated, whether the electrons are in the fast-cooling regime, and where the dissipation region is located. Therefore, we further divide our sample using $Y = 1$ as the threshold.

For FSRQs with $Y \leq 1$ (6 out of 52 FSRQs), no correlation is found between $\log Y$ and $\log \nu_{syn}^{peak}$ in either the observer's frame ($p = 0.09$) or the comoving frame ($p = 0.32$). Similarly, no correlation is found between $\log L_{bol}$ and $\log \nu_{syn}^{peak}$ in either observer's frame ($p = 0.45$) or the comoving frame ($p = 0.12$). For FSRQs with $Y > 1$ (46 out of 52 FSRQs), no correlation is found between $\log Y$ and $\log \nu_{syn}^{peak}$ in either the observer's frame ($p = 0.46$) or the comoving frame ($p = 0.36$). Likewise, no correlation is found between $\log L_{bol}$ and $\log \nu_{syn}^{peak}$ in the observer's frame ($p = 0.58$) and in the comoving frame ($p = 0.09$). These results are generally consistent with Wan et al. (2024), except for the correlations between $\log L_{bol}$ and $\log \nu_{syn}^{peak}$ in the observer's frame for $Y > 1$. Notably, when a correlation is present, the slope of the best linear fit agrees closely with Wan et al. (2024), suggesting that the dissipation region of these FSRQs is located in an accelerating jet beyond the external photon field, operating in the slow-cooling regime (Wan et al. 2024).

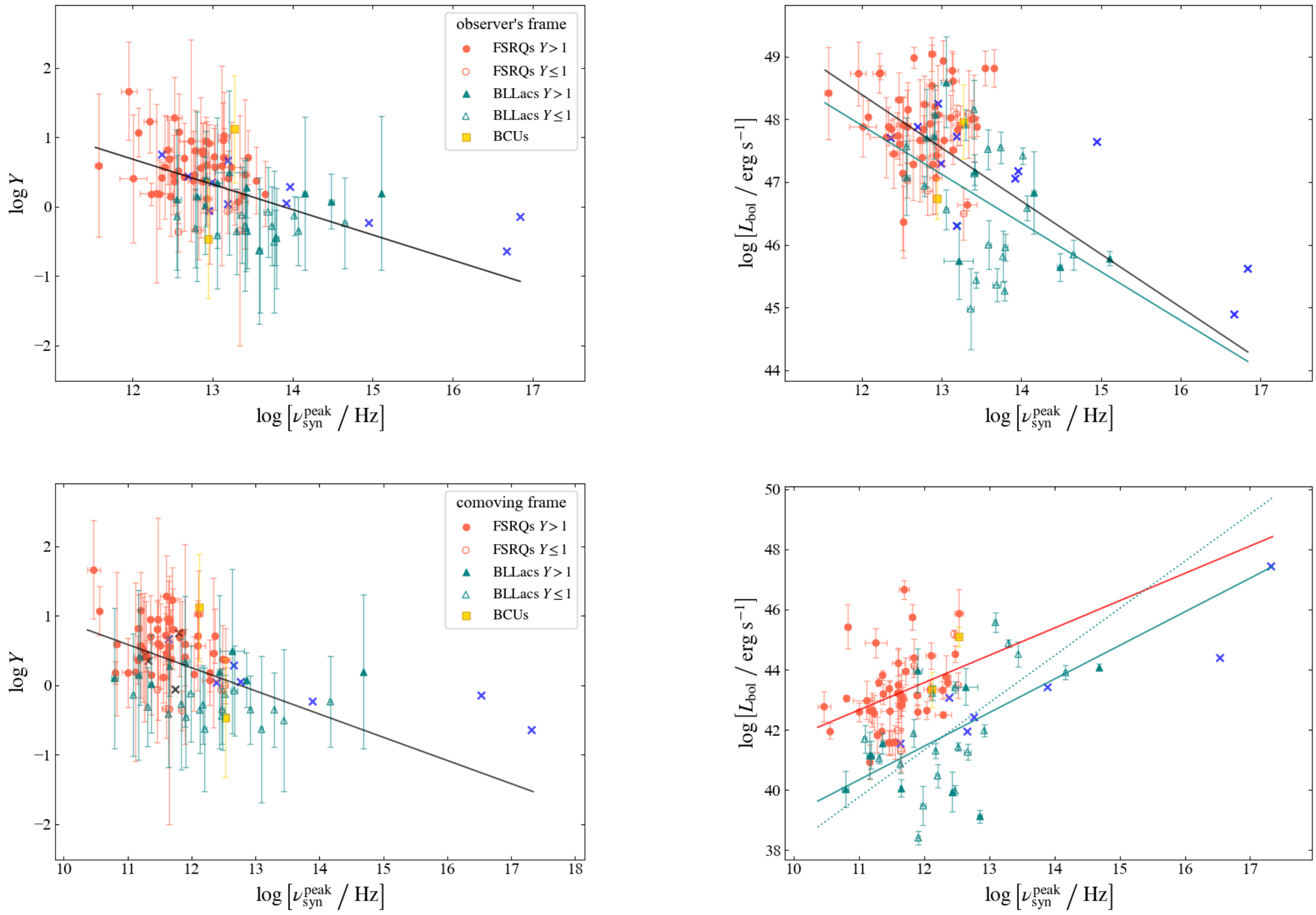

**Figure 6.** Correlations between $\log Y$ and $\log \nu_{\rm syn}^{\rm peak}$, as well as between $\log L_{\rm bol}$ and $\log \nu_{\rm syn}^{\rm peak}$, in the observer's frame (upper panels) and comoving frame (lower panels). As indicated in the inset legends, red symbols denote FSRQs and teal symbols denote BL Lacs; filled symbols represent sources with $Y > 1$, while open symbols represent sources with $Y \le 1$. The blue crosses indicate the source of unphysical curvature resulting from sparse data points. If a specific (sub-)sample exhibits a statistically significant correlation, the corresponding best-fitting linear relation is shown. The solid black line indicates the fit for the entire sample. For FSRQs, the solid, dashed, and dotted red lines represent the fits for the full sample, the $Y > 1$ subsample, and the $Y \le 1$ subsample, respectively. For BL Lacs, the solid, dashed, and dotted teal lines represent the corresponding fits same as the FSRQs.

For BL Lacs with $Y \le 1$ (18 out of 28 BL Lacs), no correlation is found between $\log Y$ and $\log \nu_{\rm syn}^{\rm peak}$ in either the observer's frame ($p$ = 0.92) or the comoving frame ($p$ = 0.90). Similarly, no correlation is found between $\log L_{\rm bol}$ and $\log \nu_{\rm syn}^{\rm peak}$ in the observer's frame ($p$ = 0.30), whereas a significant positive correlation is present in the comoving frame ($p$ = 0.004, $r$ = 0.66, $s$ = 1.57 ± 0.56). For BL Lacs with $Y > 1$ (10 out of 28 BL Lacs), no correlation is found between $\log Y$ and $\log \nu_{\rm syn}^{\rm peak}$ in either the observer's frame ($p$ = 0.98) or the comoving frame ($p$ = 0.86). Likewise, no correlation is found between $\log L_{\rm bol}$ and $\log \nu_{\rm syn}^{\rm peak}$ in either the comoving frame ($p$ = 0.06), or the observer's frame ($p$ = 0.48). These results differ from those reported in (Wan et al. 2024), except for the lack of correlation between $\log Y$ and $\log \nu_{\rm syn}^{\rm peak}$ with $Y > 1$ for BL Lacs in both frames, as well as the lack of correlation between $\log L_{\rm bol}$ and $\log \nu_{\rm syn}^{\rm peak}$ with both $Y > 1$ and $Y \le 1$ for BL Lacs in the observer's frame. Given the small sample sizes for both $Y \le 1$ and $Y > 1$, these correlations may not have strong statistical significance. A larger BL Lac sample, constructed from quasi-simultaneous multi-wavelength observations, will be necessary to confirm these trends.

### 4.5 Estimation of the Doppler Factor

As suggested by Hu et al. (2024), measuring the time lag in variations across different $\gamma$-ray energy bands offers a method to estimate the Doppler factor within the framework of the one-zone leptonic model. Relativistic electrons with higher energies (producing 1–300 GeV radiation) lose energy faster than electrons with lower energies (producing 0.1–1 GeV radiation) when the dust torus (DT) photons are dominant in the EC scattering (EC-DT), resulting in shorter cooling time-scale. So there has a time delay ($\Delta T^{\rm obs}$) between observed flux variation in the 0.1–1 GeV energy band and in the 1–300 GeV band. The difference in cooling time-scales is directly related to the energy density of the DT radiation field ($U_{\rm DT}^{\rm AGN}$) and the Doppler factor ($\delta$). The Doppler Factor can be analytically expressed as (Hu et al. 2024)

$$\delta^{\rm H} \approx 10.6 \left(\frac{1\ \rm day}{\Delta T^{\rm obs}}\right)^{1/2} \left(\frac{4.5\times10^{-5}\rm erg\ cm^{-3}}{U_{\rm DT}^{\rm AGN}}\right)^{1/2}, \quad (10)$$

where $U_{\rm DT}^{\rm AGN} \approx 4.5\times10^{-5}\rm erg\ cm^{-3}\ /\ [1 + (r^{\rm AGN}/r_{\rm DT}^{\rm AGN})^4]$ (Hayashida et al. 2012). This approach provides a model-independent estimation of the Doppler factor anchored to the $\gamma$-ray dissipation re-

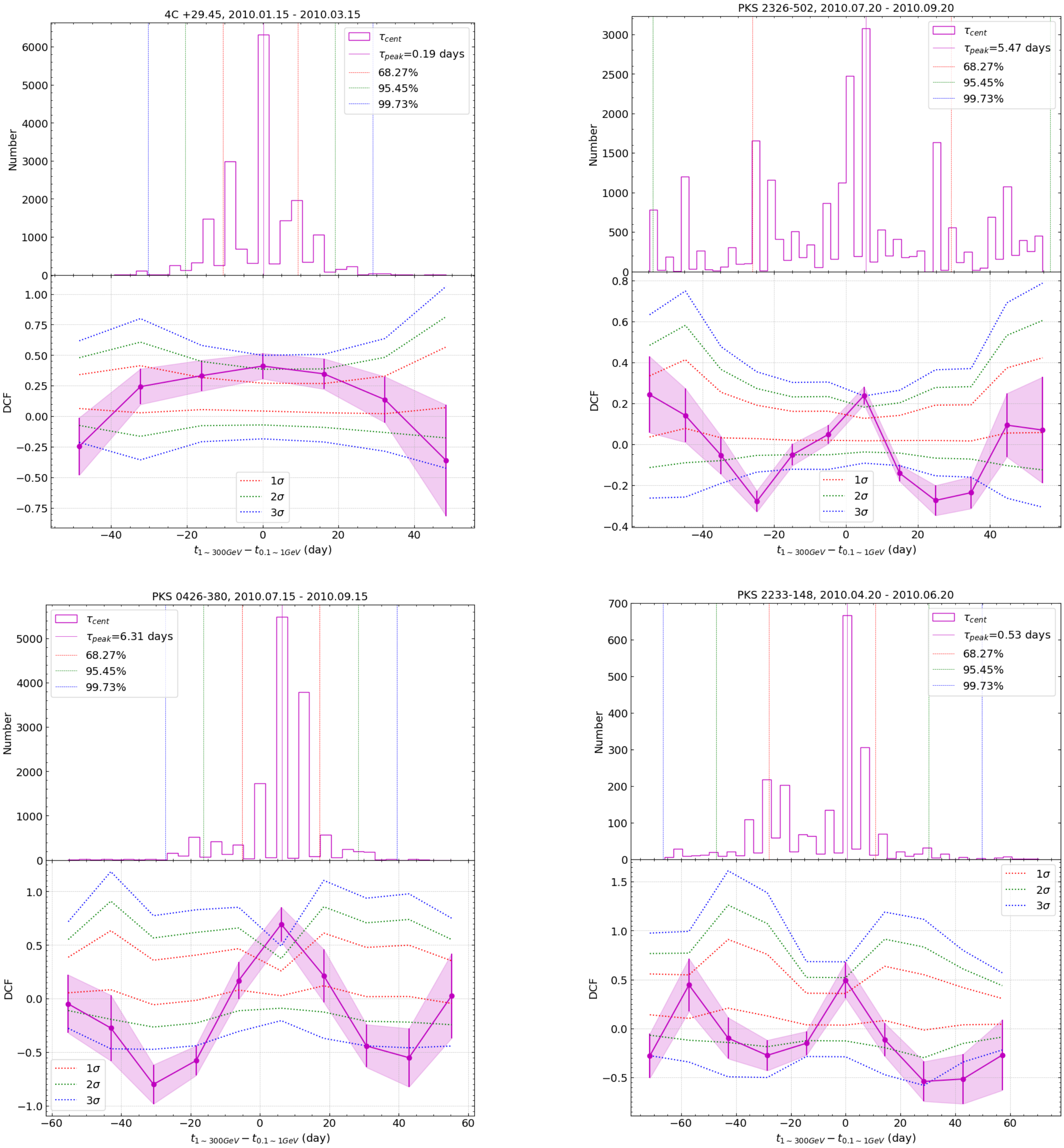

**Figure 7.** The cross-correlation results of the light curves between 0.1–1 GeV and 1–300 GeV bands for the four sources. The upper panel of each subplot shows the distribution of the central lag values obtained from the cross-correlation analysis. The histogram displays the distribution of $\tau_{\rm cent}$, the centroid lag time computed from DCF values above 60% of the peak correlation in each bootstrap resampling. The purple solid vertical line marks the peak of this distribution ($\tau_{\rm peak}$). The confidence limits of $\tau_{\rm peak}$ are indicated by red, green, and blue dashed lines. The lower panel of each subplot presents the significance of the correlation at different lags, at the $1\sigma$ (red dashed line), $2\sigma$ (green dashed line), and $3\sigma$ (blue dashed line) levels.

**Table 6.** List of Compiled Doppler Factors

| **Fermi-LAT name** (1) | **sourse name** (2) | **classification** (3) | $\Delta T^{\rm obs}$ (4) | $\delta^{\rm H}$ | $\delta^{\rm Z_1}$ | $\delta^{\rm Z_2}$ | $\delta^{\rm L}$ | $\delta^{\rm C}$ |
|---|---|---|---|---|---|---|---|---|
| 4FGL J1159.5+2914 | 4C +29.45 | FSRQ | 0.19 | 24.32 | 15.73 | 17.00 | 32.92 | 29.3 |
| 4FGL J2329.3-4955 | PKS 2326-502 | FSRQ | 5.47 | 4.53 | - | - | - | 12.0 |
| 4FGL J0428.6-3756 | PKS 0426-380 | BLL | 6.31 | 4.22 | 53.57 | 56.18 | - | 14.3 |
| 4FGL J2236.5-1433 | PKS 2233-148 | BLL | 0.53 | 14.56 | - | - | - | 59.5 |

**Note.** Columns from left to right: (1) Fermi-LAT name; (2) source name; (3) classification; (4) the time delay between observed flux enhancement in the 0.1–1 GeV energy band and the flux enhancement in the 1–300 GeV band in units of day; (5) the lower limits of Doppler factor derived from this work; (6) the Doppler factor from Zhang et al. (2020); (7) the Doppler factor calculate based on method by Zhang et al. (2020) but use the data acquired in this work. (8) the Doppler factor from Liodakis et al. (2018) (9) the Doppler factor from Chen (2018).

gion. It serves as a physical constraint for future SED modeling, reducing the degeneracy of free parameters. On the other hand, if photons from the broad-line region (BLR) dominate the EC process (EC-BLR), the significant KN effect will prolong the cooling timescale of higher-energy electrons, potentially delaying the 1–300 GeV variability or causing the enhancement in the two energy bands to occur nearly simultaneously.

From the 13 sources (10 FSRQs, 3 BL Lacs) in our sample that exhibit variability (see Table 1 for details), we selected the low-synchrotron-peaked (LSP) sources, as their high-energy peaks are likely to originate from EC radiation. Please note that for LSP BL Lacs (LBLs), despite the absence of strong emission lines, the study by Hu et al. (2024) showed that the high-energy radiation of LBLs often requires EC emission to account for their high-energy peaks. Following Hu et al. (2024), if one intends to fit the SEDs of LBLs using a one-zone SSC model, the magnetic field $B$ and $\delta$ can be estimated through (Hu et al. 2024)

$$\delta = \frac{2.8\times10^6}{1+z}\left[\frac{2f}{cR^2}\right]^{1/2}\frac{\nu_{\rm IC}^{\rm peak}}{(\nu_{\rm syn}^{\rm peak})^2}\left[\frac{L_{\rm IC}^{\rm peak}}{(L_{\rm syn}^{\rm peak})^2}\right]^{-1/2}, \quad (11)$$

$$B = \left(\frac{1+z}{2.8\times10^6}\right)^2\left[\frac{cR^2}{2f}\right]^{1/2}\left[\frac{\nu_{\rm IC}^{\rm peak}}{(\nu_{\rm syn}^{\rm peak})^2}\right]^{-2}\left[\frac{L_{\rm IC}^{\rm peak}}{(L_{\rm syn}^{\rm peak})^2}\right]^{1/2}, \quad (12)$$

where $R \approx c\tau_{\rm var}\delta$, and $f = 3$ as suggested by Hu et al. (2024). Substituting the peak frequencies and luminosities, and obtained variability timescale $\tau_{\rm var}$ of each LBL, we find $\delta \gg 50$ or $B \ll 0.01$ G which deviate significantly from the value suggested by observations, i.e., $1 \lesssim \delta \lesssim 30$ (Hovatta et al. 2009), and $0.1\ {\rm G} \lesssim {\rm B} \lesssim 10\ {\rm G}$ (O'Sullivan & Gabuzda 2009). Therefore, it is suggested that external photon fields may play a crucial role in modeling the high-energy bump of LBLs, i.e., LBLs in our sample could be masquerading BL Lacs. For the selected LSPs, we generate light curves in the 0.1–1 GeV and 1–300 GeV bands from the same two-month integration window used for the quasi-simultaneous SED construction. These sub-band light curves are used to investigate their possible correlation and time lag. We employ the discrete correlation function (DCF) method (PyDCF[7], Robertson et al. 2015) combined with a bootstrap Monte Carlo approach (Ezhikode et al. 2022) to analyze the time lag and correlation between the 0.1–1 GeV and 1–300 GeV energy bands light curves (Xiao et al. 2024a). Firstly, the DCF is computed for the original data over a time-lag range of ±60 days to obtain a preliminary cross-correlation plot. Then, a statistical sample is constructed through 50,000 bootstrap resampling iterations: each simulation randomly selects 80% of the original data pairs and recalculates the DCF using flexible time-delay bin widths (1–15 days) to determine the most promising DCF distribution. Based on these simulations, we derive confidence intervals (68.27%, 95.45%, and 99.73%) for the peak DCF value and the centroid lag (the average lag calculated from the region above 60% of the DCF peak), as shown in the upper panel of Figure 7. At last, to assess the statistical significance of the correlation and time-lag signals, we compare the original DCF curve with the confidence bands ($1\sigma$, $2\sigma$, $3\sigma$) after subtracting the original DCF value for each lag. We find that for several sources in our sample, the flux enhancement in the 1–300 GeV energy band precedes that in the 0.1–1 GeV band. Among these, we focus on four sources (4C +29.45, PKS 2326-502, PKS 0426-380, and PKS 2233-148) that exhibit the most prominent correlation signals, initially identified to be at or near the $\sim 3\sigma$ significance level. However, only PKS 0426-380 strictly exceeds the $3\sigma$ threshold. For the other three sources, the correlation is relatively marginal, with the significance reaching the $\sim 3\sigma$ only when considering their upper uncertainties. The low significance is primarily attributed to the relatively low variability level and limited photon statistics (low TS) within the selected two-month integration window, suggesting that these sources were not in a state of intense flaring. The DCF analysis results and the corresponding time lags of these four sources are given in Figure 7 and the fourth column of Table 6. Although the obtained non-zero time lags are tentative due to the marginal significance, they may still provide an independent, albeit conservative, constraint on the Doppler factor. For others, the time lag relationship is reversed (e.g., 3C 279) or show no significant difference. It implies that their EC processes might be dominated by EC-BLR, meaning the corresponding radiation zone is probably located near or with the BLR. For the four sources where the 1–300 GeV flux enhancement occur earlier, we could estimate the Doppler factor using Eq. (10). Some previous studies have suggested that the emission region lies outside the broad-line region (BLR) but within the DT (Costamante et al. 2018; Tan et al. 2020). In such case, $U_{\rm DT}^{\rm AGN} \approx 4.5\times10^{-5}{\rm erg\ cm^{-3}}$ can be treated as approximately constant. However, the emission region could also be located beyond the DT, where $U_{\rm DT}^{\rm AGN}$ would decrease sharply and the Doppler factor correspondingly increase. Thus, adopting $U_{\rm DT}^{\rm AGN} \approx 4.5\times10^{-5}{\rm erg\ cm^{-3}}$ implies that the derived Doppler factor should be regarded as a con-servative lower bound for the $\gamma$-ray emission region. The derived results are given in Table 6.

[7] https://github.com/astronomerdamo/pydcf

Zhang et al. (2020) proposed a method for estimating the Doppler factor based on correlation studies (the corresponding Doppler factor is represented by $\delta^{Z_1}$ in Table 6). By treating the $\gamma$-ray luminosity ($\log L_\gamma$) as the jet bolometric luminosity (i.e., $\log L_{\rm bol}$), the Doppler factor can be estimated using $\log L_\gamma$ and $\log L_{\rm BLR}$. Specifically, for FSRQs, $\log\delta = 0.5(\log L_\gamma - 1.178\log L_{\rm BLR} + 8.004)$; for BL Lacs, $\log\delta = 0.5(\log L_\gamma - 0.87\log L_{\rm BLR} - 6.266)$. We also calculate the Doppler factor based on this method, using the obtained $\log L_{\rm bol}$ in our work (the Doppler factor is represented by $\delta^{Z_2}$ in Table 6). Liodakis et al. (2018) and Chen (2018) also provided Doppler factor estimates (the Doppler factors are represented by $\delta^{\rm L}$ and $\delta^{\rm C}$, respectively). Liodakis et al. (2018) used the variability Doppler factor method, estimating the Doppler factor from the equipartition brightness temperature based on the radio observation. Chen (2018), on the other hand, constrained the size of the emitting region by assuming a variability time-scale of $\Delta t/(1+z) \approx 1$ day for all sources within the framework of a one-zone model. We compile the Doppler factors from these works for the same sources included in our study, and the results are presented in Table 6.

We compare the obtained and collected Doppler factors. First, we find that some values of $\delta^{Z_1}$ and $\delta^{Z_2}$ are relatively close, while others exhibit significant differences. This can be attributed to variations in the location of the radiation region or to acceleration or deceleration of the jet during certain periods, or possibly also because the use of quasi-simultaneous data or historical data may strongly influence the correlation. Additionally, the differences between $\delta^{\rm H}$ and $\delta^{\rm L}$ as well as $\delta^{\rm C}$ are evident. The approach based on radio observations has limitations (Liodakis et al. 2018), as the locations of radio and $\gamma$-ray emission regions may differ and may be farther apart. So it will ultimately lead to the deviation in the estimated Doppler factors. Chen (2018) estimated the Doppler factor by assuming a specific variability time-scale for all sources. Since actual time-scales can be longer or shorter than assumed, this method of fitting averaged states may lead to the bias in Doppler factor calculations.

## 5 CONCLUSIONS

In this work, we present a comprehensive study of quasi-simultaneous broadband SEDs for a sample of 93 blazars, including 56 FSRQs, 35 BL Lacs and 2 BCUs, from the *Fermi*-LAT Fourth Data Release. By employing cubic functions to fit the double-bump of SEDs, we derive key observational parameters ($\log\nu^{\rm peak}_{\rm syn/IC}$, $\kappa^{\rm peak}_{\rm syn/IC}$, $\log L^{\rm peak}_{\rm syn/IC}$, $\log L_{\rm bol}$). We investigate the correlations among these parameters, along with $\log M$, $\Gamma$, and $\log L_{\rm BLR}$ collected from previous works. Furthermore, we estimate the Doppler factor using the variability time lag method and discuss the physical properties of dissipation region from the implication of blazar sequence. Our main results are summarized as follows:

(1) We find several statistically significant correlations ($|r| > 0.5$, $p < 0.05$) among derived parameters. For the whole sample, the results are listed below: $\log\nu^{\rm peak}_{\rm syn}$ vs. $\log L^{\rm peak}_{\rm syn}$, $\log L_{\rm bol}$, $\log M$; $\kappa^{\rm peak}_{\rm IC}$ vs. $\log L^{\rm peak}_{\rm IC}$; $\log L^{\rm peak}_{\rm syn}$ vs. $\log L^{\rm peak}_{\rm IC}$, $\log L_{\rm bol}$, $\log M$; $\log L^{\rm peak}_{\rm IC}$ vs. $\log L_{\rm bol}$; $\log M$ vs. $\log L_{\rm BLR}$; $\log L_{\rm BLR}$ vs. . Similar correlation analyses are also performed separately for the FSRQs and BL Lacs. For FSRQs, significant correlations including: $\log\nu^{\rm peak}_{\rm syn}$ vs. $\log\nu^{\rm peak}_{\rm syn}$, $\log M$; $\log\nu^{\rm peak}_{\rm IC}$ vs. $\log L_{\rm BLR}$; $\kappa^{\rm peak}_{\rm IC}$ vs. $\log L^{\rm peak}_{\rm IC}$; $\log L^{\rm peak}_{\rm syn}$ vs. $\log L^{\rm peak}_{\rm IC}$, $\log L_{\rm bol}$, $\log M$; $\log L^{\rm peak}_{\rm IC}$ vs. $\log L_{\rm bol}$; $\log M$ vs. $\log L_{\rm BLR}$. For BL Lacs, significant correlations including: $\log\nu^{\rm peak}_{\rm syn}$ vs. $\log L^{\rm peak}_{\rm syn}$; $\log\nu^{\rm peak}_{\rm IC}$ vs. $\kappa^{\rm peak}_{\rm IC}$; $\log L^{\rm peak}_{\rm syn}$ vs. $\log L^{\rm peak}_{\rm IC}$, $\log L_{\rm bol}$; $\log L^{\rm peak}_{\rm IC}$ vs. $\log L_{\rm bol}$. The best linear fitting equations are provided for all significant correlations.

(2) We analyze correlations between peak frequencies and spectral curvatures to infer particle acceleration mechanisms in blazar jets. For BL Lacs, a strong positive correlation emerges between $2/\kappa^{\rm peak}_{\rm IC}$ and $\log\nu^{\rm peak}_{\rm IC}$ ($p = 0.003, r = 0.64, \beta = 6.04\pm1.43$). And we don't find other correlations. Our work reveals inconsistencies comparing with previous studies using different samples. And we compare these slopes ($\beta$) to theoretical models (EDPA, FFGA, stochastic acceleration). The absence of synchrotron peak-curvature correlation suggests stochastic acceleration is not the dominant mechanism. The lack of IC peak-curvature correlation in FSRQs likely stems from combined SSC/EC components and differing scattering conditions.

(3) The quasi-simultaneous analysis confirms the blazar sequence in the observers' frame, showing significant negative correlations between $\log L_{\rm bol}$ and $\log\nu^{\rm peak}_{\rm syn}$ for the whole sample ($p < 0.0001$, $r = -0.49$), alongside a moderate negative correlation between $\log Y$ and $\log\nu^{\rm peak}_{\rm syn}$ ($p < 0.0001$, $r = -0.41$). When transformed to the comoving frame using Doppler factors, though no significant correlation is found for the full sample ($p = 0.06$), significant correlations with similar slopes are detected in the FSRQ and BL Lac subclasses. Subclass analysis shows distinct behaviors: For FSRQs, there is no correlations between $\log L_{\rm bol}$ and $\log\nu^{\rm peak}_{\rm syn}$ in observers' frame but has a strong positive correlation in comoving frame. While BL Lacs show weak negative correlation in observers' frame and positive correlation in comoving frame for the same parameters. The Compton dominance correlations ($\log(Y)$ vs. $\log(\nu^{\rm peak}_{\rm syn})$) are generally weaker or absent in both frames for subclasses, particularly when subdivided by $Y \le 1$ or $Y > 1$.

(4) Based on GeV band variability time lags ($\Delta T^{\rm obs}$), the jet Doppler factor ($\delta^{\rm H}$) for four sources is estimated, with values of 4.22, 4.53, 14.56, and 24.32. Compared to the correlation method ($\delta^{\rm Z}$), the radio method ($\delta^{\rm L}$), and the time-scale assumption method ($\delta^{\rm C}$), this new approach is more closely aligned with the physics of the $\gamma$-ray radiation zone. However, the differences between these methods reflect complexities in the jet structure or acceleration processes.

## ACKNOWLEDGEMENTS

We thank the anonymous referee for insightful comments and constructive suggestions. This work is supported by the National Key R&D Program of China (2023YFB4503300), the National Natural Science Foundation of China (NSFC) under the grants No. 12203043, No. 12203024, No. 12473020, No. 12203034, No. 12473042, and No. 12373109, the Yunnan Province Youth Top Talent Project (Grant No. YNWR-QNBJ-2020-116), the CAS "Light of West China" Program, the Shanghai Science and Technology Fund (22YF1431500), the science research grants from the China Manned Space Project (CMS-CSST-2025-A07), and the Shanghai Municipal Education Commission regarding artificial intelligence empowered research.

## DATA AVAILABILITY

The data underlying this article will be shared on reasonable request to the corresponding author.

This paper has been typeset from a TEX/LATEX file prepared by the author.